\documentclass[twocolumn]{aastex63}

\usepackage{graphicx}
\usepackage[T1]{fontenc}
\usepackage{aecompl}
\usepackage[]{amsmath}
\usepackage{color}
\usepackage{xspace}
\usepackage{soul}

\usepackage{lineno}

\newcommand{\msun}{\mathrm{M}_\odot}
\graphicspath{{figures/}}

\newcommand{\ud}{\mathrm{d}}

\def\lsim{ \lower .75ex \hbox{$\sim$} \llap{\raise .27ex \hbox{$<$}} }

\def\vxm{\overline{v_{\mathrm{x',i}}}}
\def\vym{\overline{v_{\mathrm{y',i}}}}
\def\vzm{\overline{v_{\mathrm{z',i}}}}

\def\vsqxm{\overline{v^2_{\mathrm{x',i}}}}
\def\vsqym{\overline{v^2_{\mathrm{y',i}}}}
\def\vsqzm{\overline{v^2_{\mathrm{z',i}}}}

\def\vsqxym{\overline{v^2_{\mathrm{x'y',i}}}}
\def\vsqxzm{\overline{v^2_{\mathrm{x'z',i}}}}
\def\vsqyzm{\overline{v^2_{\mathrm{y'z',i}}}}

\submitjournal{ApJ}

\shorttitle{Effect of binaries on dynamical modeling with \textsc{Auriga}}
\shortauthors{Wang et al.}
\graphicspath{{./}{figures/}}

\begin{document}

\title{Unraveling the Complexity of Dwarf Galaxy Dynamics: A study of Binary Orbital Motions}

\correspondingauthor{Wenting Wang}
\email{wenting.wang@sjtu.edu.cn}


\author[0000-0002-5762-7571]{Wenting Wang}
\affiliation{Department of Astronomy, Shanghai Jiao Tong University, Shanghai 200240, China}
\affiliation{Shanghai Key Laboratory for Particle Physics and Cosmology, Shanghai 200240, China}
\author{Ling Zhu}
\affiliation{Shanghai Astronomical Observatory, Chinese Academy of Sciences, 80 Nandan Road, Shanghai 200030, China}
\author{Yipeng Jing}
\affiliation{Department of Astronomy, Shanghai Jiao Tong University, Shanghai 200240, China}
\affiliation{Shanghai Key Laboratory for Particle Physics and Cosmology, Shanghai 200240, China}
\author{Robert J. J. Grand}
\affiliation{Astrophysics Research Institute, Liverpool John Moores University, 146 Brownlow Hill, Liverpool, L3 5RF, UK}
\affiliation{Instituto de Astrof\'isica de Canarias, Calle Vía L\'actea s/n, E-38205 La Laguna, Tenerife, Spain}
\affiliation{Departamento de Astrof\'isica, Universidad de La Laguna, Av. del Astrof\'isico Francisco S\'anchez s/n, E-38206, La aguna, Tenerife, Spain}
\author{Zhaozhou Li}
\affiliation{Centre for Astrophysics and Planetary Science, Racah Institute of Physics, The Hebrew University, Jerusalem 91904, Israel}
\affiliation{Department of Astronomy, Shanghai Jiao Tong University, Shanghai 200240, China}
\affiliation{Shanghai Key Laboratory for Particle Physics and Cosmology, Shanghai 200240, China}
\author{Xiaoting Fu}
\affiliation{Purple Mountain Observatory, Chinese Academy of Sciences, 10 Yuanhua Road, Nanjing 210023, China}
\author{Lu Li}
\affiliation{Shanghai Astronomical Observatory, Chinese Academy of Sciences, 80 Nandan Road, Shanghai 200030, China}
\author{Jiaxin Han}
\affiliation{Department of Astronomy, Shanghai Jiao Tong University, Shanghai 200240, China}
\affiliation{Shanghai Key Laboratory for Particle Physics and Cosmology, Shanghai 200240, China}
\author{Ting S. Li}
\affiliation{David A. Dunlap Department of Astronomy \& Astrophysics, University of Toronto, 50 St. George Street, Toronto, ON, M5S3H4 Canada}
\affiliation{Dunlap Institute for Astronomy \& Astrophysics, University of Toronto, 50 St George Street, Toronto, ON M5S 3H4, Canada}
\author{Fabo Feng}
\affiliation{Tsung-Dao Lee Institute, Shanghai Jiao Tong University, Shengrong Road 520, Shanghai, 201210, China}
\affiliation{School of Physics and Astronomy, Shanghai Jiao Tong University, 800 Dongchuan Road, Shanghai 200240, China}
\author{Carlos Frenk}
\affiliation{Institute for Computational Cosmology, Department of Physics, Durham University, South Road, Durham DH1 3LE, UK}




\begin{abstract}

We investigate the impact of binary orbital motions on the dynamical modeling of dwarf galaxies with intrinsic line-of-sight velocity dispersions ($\sigma_{v_r}$) of 1 to 9~km/s. Using dwarf galaxies from the \textsc{auriga} level-2 and level-3 simulations, we apply the Jeans Anisotropic Multi-Gaussian Expansion modelling to tracer stars before and after including binaries to recover the dynamical masses. The recovered total masses within the half-mass radius of tracers, $M(<r_\mathrm{half})$, are always inflated due to binary motions, with greater inflations occurring for smaller $\sigma_{v_r}$. However, many dwarf galaxies experience central density {\it deflated} due to binary motions, with little dependences on $\sigma_{v_r}$. This is due to the negative radial gradients in the velocity dispersion profiles, with the fractional inflation in $\sigma_{v_r}$ due to binaries more significant in outskirts. 
An extreme binary fraction of 70\% can lead to central density deflation of up to 10-20\% at 3~km/s~$<\sigma_{v_r}<$8~km/s, with $M(<r_\mathrm{half})$ inflated by 4\% at 9~km/s and up to 15\% at 3~km/s. A lower binary fraction of 36\% leads to similar deflations, with the inflations decreasing to approximately 10\% at 3~km/s and becoming statistically insignificant. The choice of binary orbit distribution models does not result in significant differences, and observational errors tend to slightly weaken the deflations in the recovered central density. Two observations separated by one year to exclude binaries lead to almost zero inflations/deflations for a binary fraction of 36\% over 3~km/s$<\sigma_{v_r}<$9~km/s. For $\sigma_{v_r}\sim$1~km/s to 3~km/s, a binary fraction of 70\% (36\%) still results in 60\% (30\%) to 10\% (1\%) of inflations in $M(<r_\mathrm{half})$, even with two-epoch observation.



\end{abstract}



\section{Introduction}
\label{sec:intro}

Understanding the dark matter content and inner density profiles of dwarf galaxies has long been a hotly debated topic in galaxy formation and cosmology. For example, the so-called ``core-cusp'' problem has been raised to the standard theory, in the way that dark matter only simulations predict inner density slopes close to $-1$ (cusp), whereas the modeling of gas rotation curves or stellar kinematics in the central regions of low surface brightness galaxies, gas rich dwarfs and dwarf spheroids favor inner slopes close to $0$ (core), which brings in tension with the theory 
\citep[e.g.][]{1994ApJ...427L...1F,1994Natur.370..629M,2001AJ....122.2396D,2004MNRAS.351..903G,
2010AdAst2010E...5D,2017ARA&A..55..343B}. For another instance, the discoveries of dwarf galaxies with very low fractions or no dark matter has invoked puzzles to the theory as well, because dwarf galaxies are believed to be typically dark matter dominated systems \citep[e.g.][]{2008ApJ...688.1009M,2009ApJ...705..758M,2016MNRAS.459.2370T,2019MNRAS.488.2743T,2020MNRAS.491.3496C,2021MNRAS.505.5686C}.

Observationally, constraints on the dark matter content of dwarf galaxies are usually based on dynamical modeling of observed member stars or gas rotation curves, while it has been pointed out that invalid assumptions behind various types of dynamical models may cause wrong conclusions, especially for individual or small sample systems \citep[e.g.][]{2018MNRAS.474.1398G,2022ApJ...941..108W}. In addition to the classical Milky Way (MW) dwarfs, a large number of dwarf spheroidal satellite galaxies have been discovered around our MW. These objects are much fainter, having larger mass-to-light ratios and velocity dispersions in the range of $<\sim$3~km/s to 7-8~km/s. Modeling of these fainter dwarf galaxies is even more susceptible to errors than more massive MW classical dwarf galaxies. 

The sources of errors not only include statistical ones, because such faint dwarf galaxies usually have a small number of member stars observed, but also, more importantly, include systematic errors such as the contamination by foreground stars and inflation of the velocity dispersion due to orbital motions of binary stars. The binary orbital motion is the most difficult to be corrected, which not only depends on a correct estimate of binary fraction in the tracer star sample based on multi-epoch observations of the line-of-sight velocities (LOSV or LOSVs), but also requires a good knowledge of the binary orbital element distributions. 

With orbital element distributions constrained from solar neighborhood stars, dwarf spheroids with velocity dispersions of 4 to 10~km/s are claimed unlikely to have their intrinsic dispersions significantly inflated by more than 20-30\% \citep[e.g.][]{1996MNRAS.279..108H,2010ApJ...721.1142M}. On the other hand, since the typical velocities of binary orbital motions can reach a few km/s, dwarf systems with intrinsic dispersions $<\sim$4~km/s have a higher risk to have their velocity dispersions more significantly boosted \citep[e.g.][]{2007ApJ...670..313S,2007MNRAS.380..281M,2010ApJ...722L.209M,2011ApJ...736..146K,2017AJ....153..254S,2019MNRAS.487.2961M,2022ApJ...939....3P}. Note in some of the previous studies, 3-$\sigma$ clippings of extreme velocities are applied, and sometimes multi-epoch data are used to exclude binaries.

Most of the previous analysis, however, were based on Monte Carlos simulations with Gaussian intrinsic velocity dispersions. Real dwarf galaxies can have negative radial gradients in their velocity dispersion profiles, with higher dispersions in the very center, and lower dispersions in outskirts. For example, prominent negative radial gradients were seen in a few classical dwarf spheroids, such as in the metal rich population of Sculpter \citep[e.g.][]{2008ApJ...681L..13B,2016MNRAS.463.1117Z}, and in the metal poor populations of Ursa Minor \citep[e.g.][]{2020MNRAS.495.3022P} and of Fornax \citep[e.g.][]{2012ApJ...756L...2A}. For ultra faint dwarfs\footnote{Ultra faint dwarf galaxies are typically defined to have stellar masses smaller than $10^5\msun$ \citep[e.g.][]{2019ARA&A..57..375S,2021MNRAS.504.3509O}.}, the velocity dispersions of which are usually believed to be most significantly affected by binaries, their number of observed stars is currently too few to enable robust measurements of radial gradients. However, if the velocity dispersion profiles are radius dependent, the total mass enclosed within different radii would be biased by binary orbital motions to different levels. Thus investigating more realistically simulated dwarf galaxy systems could bring new insights towards how binary orbital motions affect the constraints on the underlying density profiles of dwarf galaxies. 

The resolution limit of ultra faint dwarf galaxies in modern simulations is a crucial issue. Individual stars are not resolved yet, and the star particles in nowadays hydrodynamical simulations represent single stellar populations. The average stellar masses of star particles or gas cells are often on the order of $>\sim10^{4-5}\msun$. For example, the baryonic particle mass of the IllustrisTNG50 simulation is $\sim10^4\msun$ \citep{2019MNRAS.490.3196P}. For dwarf galaxies with stellar masses of $10^{4-5}\msun$, their velocity dispersions usually range from a few to 10~km/s. Though their velocity dispersions are likely inflated by binaries \citep[e.g.][]{1996MNRAS.279..108H,2010ApJ...721.1142M}, the number of tracer star particles is too few for proper dynamical modelings. There are only a few to about ten star particles for $10^{4-5}\msun$ dwarfs in TNG50. 

The resolution is significantly higher in the cosmological zoom-in hydrodynamical simulation, \textsc{auriga} \citep{2017MNRAS.467..179G}, which are particularly aimed to resolve the evolution of MW-mass systems and their population of satellites. For example, the level-2 suite of \textsc{auriga} simulations have an average star particle mass of $\sim800\msun$. Hence for dwarfs with stellar masses of $10^{4-5}\msun$, they can have a few tens up to a few thousands of star particles, enabling the investigations on how binary orbital motions affect the mass constraints for more realistic dwarf systems in modern hydrodynamical simulations. 

In this paper, we take advantage of the \textsc{auriga} simulations. Since binaries  are not resolved, we incorporate binary orbital motions by sampling their orbital element distributions based on different models, while the original motions of star particles are treated as the motions of barycenters for binaries. We start with the so far highest resolution of level-2 simulations. Moreover, since level-2 has only one MW-like system so far and with a few tens of star particles as tracers at the low mass end, the statistical errors are still large, we will also use the lower resolution suite of level-3 simulations plus a scaling method. The \textsc{auriga} simulations and the scaling approach have enabled us to investigate the effect of binaries upon constraining the underlying density profiles, and in particular, our analysis is based on more realistic dwarf systems with radius dependent velocity dispersion profiles. The best constrained mass at different radii can be directly compared  before and after incorporating binary motions. Note, however, we do not evaluate the performance of \textsc{jam} in this paper, since the focus of this paper is the effect of binary motions. We refer the readers to \cite{2022ApJ...941..108W} (hereafter Paper I) about the performance of \textsc{jam}, in which we have performed detailed investigations on how the best fits by \textsc{jam} may deviate from the truth for dwarf systems in \textsc{auriga} simulations.

The layout of the paper is as follows. We first introduce the \textsc{auriga} suite of simulations, sample of simulated dwarf galaxy systems, mock stars and the models of binary orbital element distributions in Section~\ref{sec:data}. Our dynamical modeling approach is the axis-symmetric Jean Anisotropic multi-Gaussian expansion method (Section~\ref{sec:methods}). The model constraints and the comparison before and after including binaries will be shown in Section~\ref{sec:results}, with discussions on different binary models and binary fractions, with or without observational errors and multi-epoch observations. We discuss and conclude in the end (Section~\ref{sec:concl}).

\section{Data}
\label{sec:data}

In this section, we first introduce the \textsc{auriga} simulations. We then move on to introduce the selection of dwarf galaxies, mock stars, the incorporation of binary motions, the creation of mock galaxy images and multi-Gaussian decomposition of the stellar component. 

\subsection{The \textsc{auriga} suite of simulations}

Details about the \textsc{auriga} simulations can be found in \cite{2017MNRAS.467..179G} and \cite{2018MNRAS.481.1726G}. Here we make a brief introduction. 

The \textsc{auriga} simulations are a suite of cosmological zoom-in simulations, with the parent systems identified as those isolated and MW-mass halos from the dark matter only simulations of the EAGLE project \citep{2015MNRAS.446..521S}, the evolution of which are re-simulated with higher resolutions and traced from redshift $z=127$ to $z=0$. The cosmological parameters adopted are from the third-year Planck data \citep{2014A&A...571A..16P} with $\Omega_m=0.307$, $\Omega_\Lambda=0.693$, $\Omega_b=0.048$ and $H_0=67.77\mathrm{km s^{-1} Mpc^{-1}}$. 

The simulations were performed using the magneto-hydrodynamical code \textsc{arepo} \citep{2010MNRAS.401..791S} with full baryonic physics, which incorporates a comprehensive galaxy formation model. The physical mechanisms of the galaxy formation model include atomic and metal line cooling \citep{2013MNRAS.436.3031V}, a uniform UV background \citep{2009ApJ...703.1416F}, a sub grid model of the interstellar medium and star formation processes \citep{2003MNRAS.339..289S}, metal enrichment from supernovae and AGB stars \citep{2013MNRAS.436.3031V}, feedback from core collapse supernovae \citep{2010MNRAS.406..208O} and the growth and feedback from supermassive black holes \citep{2005MNRAS.361..776S}. A uniform magnetic field with co-moving strength of $10^{-14}$~G is set at redshift $z=127$, which quickly becomes subdominant in collapses halos \citep{2013MNRAS.432..176P,2017MNRAS.469.3185P}.

We will use the ``level-2'' and ``level-3'' resolutions of \textsc{auriga} simulations. There are six MW-like systems in the level-3 simulations, named Au6, Au16, Au21, Au23, Au24 and Au27. The virial masses\footnote{The virial mass, $M_{200}$, is defined as the mass enclosed in a radius, $R_{200}$, within which the mean matter density is 200 times the critical density of the universe.} of their host dark matter halos are in the range of 1-2$\times10^{12}\msun$, to represent the virial mass of our MW \citep[e.g.][]{2020SCPMA..6309801W,2015MNRAS.453..377W}. The typical dark matter particle mass of level-3 resolution is about $4\times10^4\msun$, while the average baryonic particle mass is about $5\times10^3\msun$. More recently, Au6 has been re-simulated with level-2 resolution. It has a baryonic mass resolution of $\sim$800$\msun$. There are approximately five times as many satellite galaxies at this high resolution compared to a standard baryonic resolution simulation of $10^{4-5}\msun$ for the same system \citep{2021MNRAS.507.4953G}. Note level-3 and level-2 resolutions will meet different purposes in this study, which will be explained in Section~\ref{sec:dwarf}.

\subsection{Dwarf galaxies}
\label{sec:dwarf}

\begin{figure*} 
\begin{center}
\includegraphics[width=0.8\textwidth]{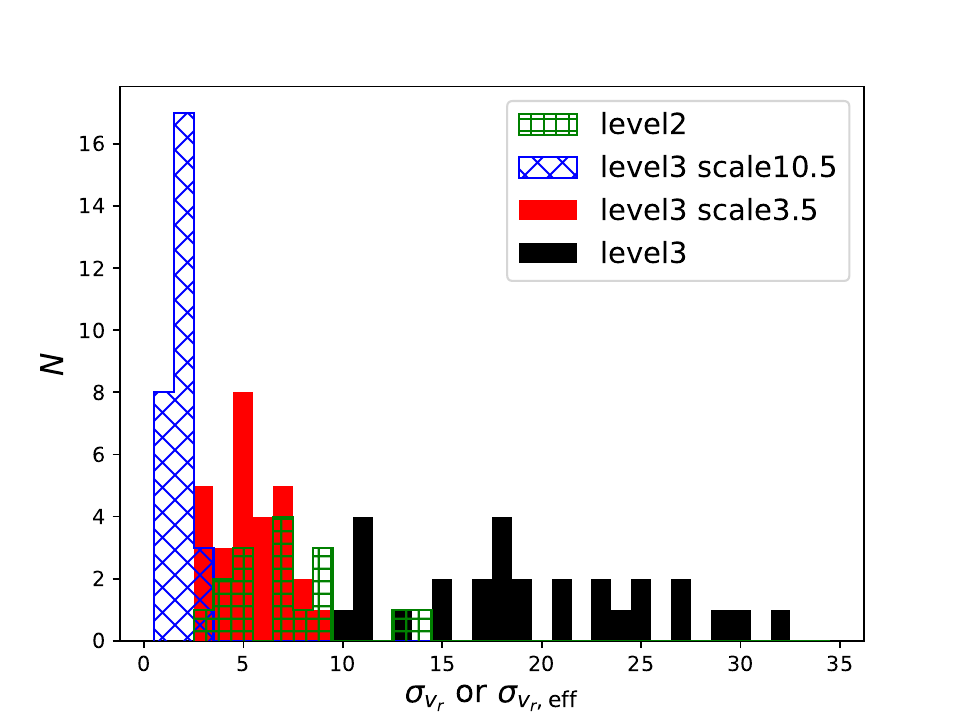}%
\end{center}
\caption{Black, red and blue histograms are the line-of-sight velocity dispersions of dwarf galaxies selected from the \textsc{auriga} level-3 simulations. The black histogram shows the original line-of-sight velocity dispersions, on the basis of which the velocity dispersions of the red and blue histograms are manually reduced by factors of 3.5 and 10.5, respectively. The original stellar masses of these level-2 systems are in the range of $10^{7.5}$ to $10^{9}\msun$. The scaling has enabled us to equivalently treat them as lower mass dwarf galaxies when investigating the effect of binary motions, while still maintain enough number of star particles as tracers. The green histogram is based on dwarf galaxies from the \textsc{auriga} level-2 simulations with stellar masses smaller than $10^{7.5}\msun$. }
\label{fig:vhist}
\end{figure*}

Each of the MW-like systems in level-3 and level-2 simulations has its dwarf satellite galaxies. For dwarfs with stellar masses of $10^{4-5}\msun$, the number of available star particles ranges from only $<\sim10$ to at most a few hundred in level-3, while the level-2 resolution can have a few tens to $>\sim$1,000 star particles as tracers. 

We first use the population of dwarf galaxies in the level-2 simulation for initial analysis. We select dwarf systems which are less massive than $10^{7.5}\msun$ in stellar mass and also have more than 40 star particles. Massive dwarfs with stellar masses much greater than $10^{5}\msun$ are unlikely to be significantly affected by binary orbital motions upon dynamical modeling, but we include them for comparisons with less massive systems. In addition, systems with minor axes strongly mis-aligned with their spin axes severely deviate from axis-symmetry and cannot be well fit by our dynamical modeling method, which are excluded from our analysis. In the end, we have 17 dwarf systems from level-2 (see Section~\ref{sec:methods} for details). 

On the other hand, although the lower limit of 40 star particles corresponds to approximately $\sim3\times10^4\msun$ in level-2 and 40 tracers enables reasonable dynamical constraints, the statistical errors in the best fits are quite large. Besides, with only a few tens of star particles, the internal dynamics of dwarf systems might not be well resolved, hence preventing us from robust investigations on systematics introduced by binaries based on level-2 resolution. We thus only use level-2 for initial analysis, and will focus on discussing results based on level-3 simulations with a scaling method. 

We select dwarf galaxies having at least 6,000 star particles from the level-3 simulations. This corresponds to a lower limit of $\sim10^{7.5}\msun$ in stellar mass. The large number of star particles to be used as dynamical tracers can suppress the size of statistical errors, but massive dwarfs with stellar masses greater than $10^{7.5}\msun$ are unlikely to have their velocity dispersions significantly boosted by binary orbital motions. Hence we will manually increase the line-of-sight velocities due to binary orbital motions (see Section~\ref{sec:binary_motion} and Equation~\ref{eqn:binarylosv}) by factors of 3.5 or 10.5 in our analysis. Relatively, this is equivalent to decreasing the velocity dispersions of massive dwarfs in level-3 by factors of 3.5 or 10.5, with respect to the level of binary motions. The scaling method enables us to effectively investigate how binary motions affect dynamical modeling outcomes for low-mass dwarfs, while at the same time maintains enough number of star particles as tracers. 

In the end, we have 28 systems from level-3, which have been used Paper I. Note we can also use massive dwarfs in level-2 with this scaling method, but there are more massive dwarfs in level-3, as it has six MW-like systems. However, we should bear in mind that the kinematics of more massive dwarfs in \textsc{auriga} might not fully represent the kinematics of less massive ultra faint dwarfs, though at the current stage the internal dynamics of ultra faint dwarfs are not as well resolved as more massive satellites. This is currently the best approach we can adopt with \textsc{auriga}, which is so far one of the highest resolution hydrodynamical simulation for the MW systems.

In Figure~\ref{fig:vhist}, we show histograms of the intrinsic LOSV dispersions for dwarf galaxies selected following the details above. Black, red and blue histograms are level-3 systems. The black histogram shows the original velocity dispersions. The velocity dispersions of the red and blue histograms are smaller than those of the black histogram by factors of 3.5 and 10.5, respectively. With this scaling method, it is equivalent to say that we are investigating systems with velocity dispersions in the range of $\sim$3 to 9~km/s (red histogram) and 1 to 3~km/s (blue histogram). Besides, the green histogram shows dwarf systems from the level-2 resolution, without any scaling to $\sigma_{v_r}$. They have velocity dispersions between 3 and 14~km/s. 

Throughout this paper, the intrinsic velocity dispersions for dwarf systems in level-3 resolution will be shown after being scaled by factors of 3.5 or 10.5, and we call it the effective LOSV dispersion or $\sigma_{v_r,\mathrm{eff}}$. Note the velocity and velocity dispersion values that we use in \textsc{jam} modeling are the actual values, rather than the scaled values. What we directly scale are the part in the LOSVs due to binaries only. On the other hand, when we present the density profiles, velocity moments and the enclosed masses within some given radius, the radial scales are still based on the original coordinates in the simulation, without any scaling. This is because it is not straight-forward to figure out a universal scaling of the coordinates, after scaling the velocity dispersions. Of course, we find correlations between the velocity dispersions of star particles in dwarf satellite galaxies and the total mass or scale radius of their host subhalos in \textsc{auriga}, but the amount of scatter in the correlation is not negligible. So we choose to present the radial scales based on the original coordinates without scalings. However, the readers may wonder what are approximately the scaling factors in the radius, given the 10.5 and 3.5 factors of scalings in velocity dispersions. For virialized dark matter halos, the virial velocity and virial radius are proportional to each other according to the spherical collapse model \citep[][]{1972ApJ...176....1G}. So the readers can approximately divide the radii by factors of 10.5 or 3.5 accordingly to have the rough numbers.

\subsection{Mock stars}
\label{sec:star}

To create mock ``observed'' stars in each dwarf, we start from the star particles in the simulation, and subtract from them the stellar mass weighted mean coordinates and velocities of all bound particles belonging to each dwarf, to eliminate perspective accelerations \citep{1961MNRAS.122..433F}. Note although each star particle is a single stellar population, we treat them as individual observed stars or unresolved binaries (see Section~\ref{sec:binary_motion} for more details), i.e., we ignore the original information such as the total stellar mass or luminosity of the star particle. We place the observer on the disk plane of the host galaxy, which is defined as the plane perpendicular to the minor axis of all bound star particles with galactocentric distances smaller than 20~kpc. The observer is 8~kpc away from the galactic center, with a random position angle.

The coordinates and velocities are then transformed to the observing frame. The $z'$-axis of the observing frame is chosen as the line-of-sight direction. The $x'$-axis (major axis) is the cross product between the spin axis of the dwarf galaxy and the $z'$-axis, which is projected on the ``sky''. The $y'$-axis (minor axis) is the cross product between $z'$ and $x'$ vector, taking minus sign, which represents a left-handed observing frame \citep{2013MNRAS.436.2598W}.

\subsection{Incorporating binary orbital motion}
\label{sec:binary_motion}

Star particles from dwarfs in the \textsc{auriga} level-3 and level-2 resolutions are used as dynamical tracers. If a star particle is determined to be a binary, its original velocity is adopted to represent the motion of the center of mass (CM), and we incorporate the binary orbital motion by sampling two orbital element distribution models based on solar neighborhood observations. The orbital elements used to determine the LOSV include the mass of the primary star, $m_1$, the mass ratio of the binaries, $q=m_2/m_1$, orbital period, $P$, orbit eccentricity, $e$, the inclination angle of the orbit plane, $i$, the argument of periapse, $\omega$, and the true anomaly, $f$. In the following, we introduce their relation to the LOSV and their meanings.
 
For $m_1$ and $m_2$ being the stellar masses of the primary and secondary stars, the LOSV of the primary star with respect to the CM is described by the following equation

\begin{equation}
v_{z'}=\frac{m_2}{m_1+m_2}\frac{2\pi a\sin{i}}{P\sqrt{1-e^2}}(\cos(\omega+f)+e\cos{\omega}),
\label{eqn:binarylosv}
\end{equation}



\begin{figure} 
\begin{center}
\includegraphics[width=0.49\textwidth]{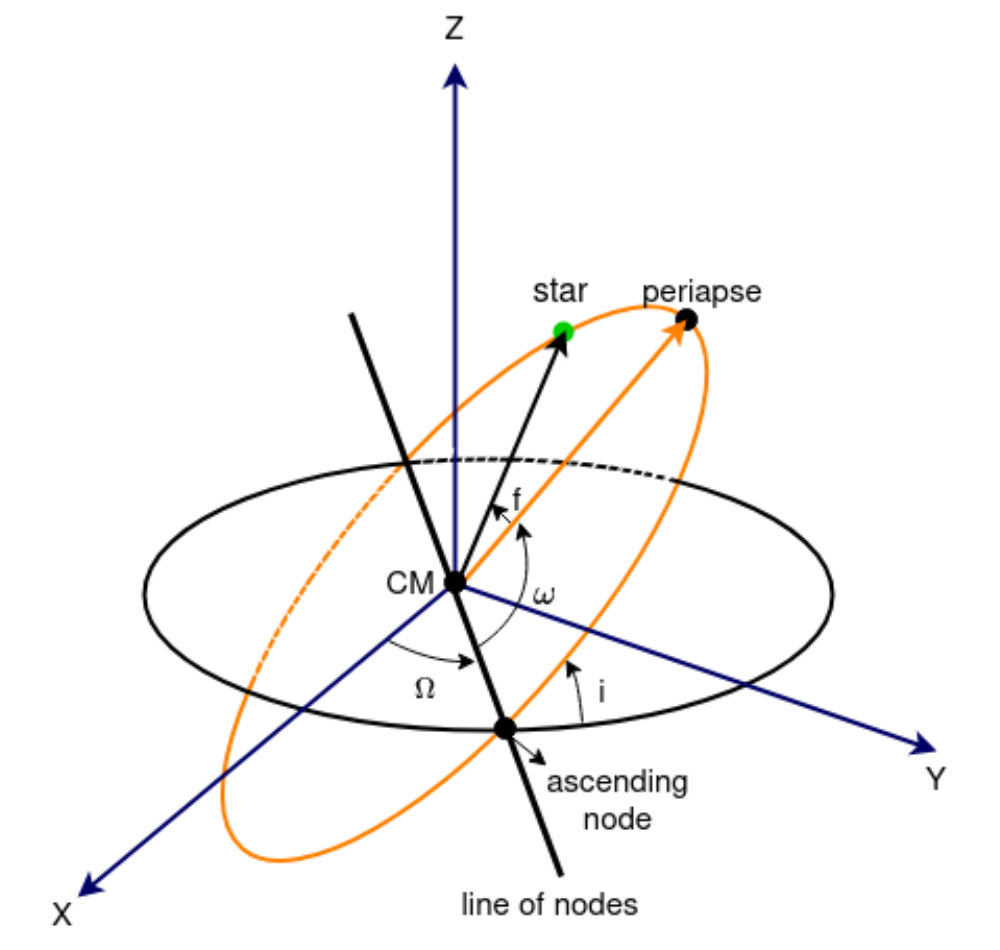}%
\end{center}
\caption{A demonstration of the the inclination angle of the binary orbit plane (orange ellipse) with respect to the reference plane (black ellipse), $i$, the argument of periapse, $\omega$, and the true anomaly, $f$. The origin is chosen as the center of mass (CM) of the binary system. The green dot represents the current position of the primary star. $\Omega$ is the angle between the reference line and the radius vector to the ascending node, which is not relevant for the line of sight velocity, with the observer put at the positive $Z$-axis. The readers can refer to Figures~4 and 7 of \cite{2010exop.book...15M} for similar versions of the figure.}
\label{fig:orbit_element}
\end{figure}

where $P$ is the orbital period. $e$ is the orbit eccentricity and $a$ is the orbit major axis.
$a$ is not independent and is linked to $P$ through $\frac{4\pi^2}{P^2}a^3=G(m_1+m_2)$. $i$ is the inclination angle between the orbit plane (the orange ellipse in Figure~\ref{fig:orbit_element}) and the reference plane (the black ellipse in Figure~\ref{fig:orbit_element}), and here the reference plane is defined to be perpendicular to the line of sight (positive $Z$-axis of Figure~\ref{fig:orbit_element}). 
The line formed by the intersection of the orbit and reference planes is called the line of nodes. The ascending node is the point in both planes where the orbit crosses the reference plane moving from below to above the plane. The angle between this same radius vector and the periapse of the orbit is called the argument of periapse, $\omega$. $f$ is the true anomaly, which is the angle between the line connecting the periastron to the CM and the CM to the star. 
The binary motions with respect to the CM are thus included through the above equation and by sampling the distributions of $m_1$, $q$, $P$, $e$, $i$, $\omega$ and $f$. 

For each star particle, we first determine whether it is a binary according to a given binary fraction ($f_\mathrm{binary}$). We try two different $f_\mathrm{binary}$, 70\% and 36\%. The fraction of 70\% might be reached in young star-forming regions, but might not be realistic for old dwarf galaxies systems, whose binaries might have been significantly dissolved by tidal disruptions. Note the MW field binary fraction is smaller than 50\% \citep[e.g.][]{2013ApJ...779..116M}. However, in a few recent studies, fractions as high as 50\% to 78\% were reported with high confidence levels for Draco, Ursa Minor and Reticulum II \citep[][]{2018AJ....156..257S,2019MNRAS.487.2961M}, and it was also shown that the binary fractions in MW dSphs can vary significantly. Thus we try $f_\mathrm{binary}$ of both 70\% and 36\%, not only to cover a wide possibility, but also to test extreme cases. 

Among the full set of orbital elements, the distribution of $i$, $\omega$ and $f$ are not model dependent. We first introduce how we sample them. The distribution of $\omega$ follows the uniform distribution over 0 and $2\pi$, and the distribution of the inclination angle, $i$, is proportional to $\sin i$ in our analysis, assuming random orientation. However, observational evidences exist to show that the Oort cloud of the outer solar system shows some alignments with the Galactic disk due to the Galactic torque \citep[e.g.][]{1987A&A...187..913D,2020AJ....160..134H,2014MNRAS.442.3653F}. If the tidal torque of Galactic disk has similar effect on binary systems in dwarf galaxies, we may expect the distribution of $i$ to be modified. Nevertheless, the outer edge of the Oort cloud is about 10,000 to even 100,000 AU from the Sun. Such a wide scale corresponds to wide binary systems with small contributions to the LOSVs \citep[e.g.][]{2021MNRAS.506.2269E,2020ApJS..246....4T}. Besides, most dwarf spheroids are far away from the Galactic disk, and thus their member stars are unlikely to be significantly affected by the disk torque. So we believe Galactic tidal torque would not significantly affect our analysis here. 

The true anomaly, $f$, depends on the eccentricity and does not have analytic solution, and it does not change linearly with time. On the other hand, the mean anomaly, $M$, scales linearly with time. The relation between $f$ and $M$ is
\
\begin{equation}
M=E-e\mathrm{sin}E,
\label{eqn:ME}
\end{equation}
\
\begin{equation}
\mathrm{sin}f=\sqrt{1-e^2}\mathrm{sin}E/(1-e\mathrm{cos}E),
\label{eqn:Ef1}
\end{equation}
and 
\begin{equation}
\mathrm{cos}f=(\mathrm{cos}E-e)/(1-e\mathrm{cos}E),
\label{eqn:Ef2}
\end{equation}
where $E$ is the eccentric anomaly. 

We first sample the mean anomaly, $M$, from a uniform distribution from 0 to $2\pi$. $E$ is then solved numerically through Equation~\ref{eqn:ME}, and in the end we solve the true anomaly, $f$, through equations~\ref{eqn:Ef1} and \ref{eqn:Ef2}. Note if uniformly sampling $f$ between 0 and $2\pi$, we end up with more epochs close to the apoastrons than it should be, which can cause the velocity dispersions to be mistakenly inflated by more than a factor of three \citep[e.g.][]{1996MNRAS.279..108H}. 

The distributions of $m_1$, $q$, $P$ and $e$ are model dependent. In all cases, we assume the primary stars are red giants. We first fix $m_1=0.8\msun$ when using the \textsc{Auriga} level-2 simulations for initial analysis. We then more carefully sample $m_1$ through the Kroupa initial mass function corrected for binaries \citep{2002Sci...295...82K}, when using \textsc{Auriga} level-3 simulations. The allowed mass range is determined through a 10~Gyr and solar metallicity Parsec stellar evolution isochrone \citep{2012MNRAS.427..127B}, and to select the part for red giants, we require the surface gravity to be smaller than $\log g=3.8$, i.e., red giants. Nevertheless, it was shown that the amount of inflations in the velocity dispersion changes very little when $m_1$ is varied over a reasonable range \citep[e.g.][]{1996MNRAS.279..108H}. 

To sample the distributions of $q$, $P$ and $e$, we adopt two different models of \cite{1991A&A...248..485D} and \cite{2017ApJS..230...15M}. The model of \cite{1991A&A...248..485D} is simple to use, and have been adopted in many studies even recent ones. On the other hand, \cite{2017ApJS..230...15M} explicitly considered joint distributions of different orbital elements through a variety of more recent observations. Many recent studies reported prominent correlations among the distributions of different orbital elements, and thus proper modeling of the joint distributions across different orbital elements is necessary \citep[see also, e.g.,][]{2019MNRAS.490..550L,2022arXiv220605595G}. 
 
Throughout this paper, we call the orbital element distribution taken from \cite{1991A&A...248..485D} as model-I, and the distribution of \cite{2017ApJS..230...15M} as model-II. Here we briefly introduce the models in the following.

\subsubsection{model-I}

\cite{1991A&A...248..485D} modeled the distribution of mass ratio between the secondary and primary stars, $q=m_2/m_1$, through the following functional form

\begin{equation}
\frac{\ud N}{\ud q}=\mathrm{exp}(-\frac{(q-\mu_q)^2}{2\sigma_q^2}), 
\end{equation}
where $\mu_q=0.23$ and $\sigma_q=0.42$. 

The orbital period distribution of \cite{1991A&A...248..485D} takes the following log-normal form

\begin{equation}
\frac{\ud N}{\ud \log P}=\mathrm{exp}(-\frac{(\log P-\mu_{\log P})^2}{2\sigma_{\log P}^2}), 
\end{equation}
where $\mu_{\log P}=4.8$ and $\sigma_{\log P}=2.3$. A similar distribution was reported by \cite{2010ApJS..190....1R} 
with $\mu_{\log P}=5.03$ and $\sigma_{\log P}=2.28$, which barely affects our results.

The distribution of eccentricity depends on the orbital period $P$ and has the following form \citep[see also][]{1996MNRAS.279..108H}
\begin{align}
 e&=0, &&P<11\mathrm{days}; \nonumber \\
 \frac{\ud N}{\ud e}&=\mathrm{exp}(\frac{-(e-0.3)^2}{0.16^2}), &&11\mathrm{days}<P<1000\mathrm{days}; \nonumber \\
 \frac{\ud N}{\ud e}&=2e, &&P>1000\mathrm{days}.
\end{align}

When adopting model-I, we try different sampling of $m_1$ and $f_\mathrm{binary}$: 1) $m_1=0.8\msun$ and $f_\mathrm{binary}$ of 70\% (model-Ia); 2) $m_1$ sampled from Kroupa initial mass function and $f_\mathrm{binary}$ of 70\% (model-Ib) and 3) $m_1$ sampled from Kroupa initial mass function and $f_\mathrm{binary}$ of 36\% (model-Ic). Note when we apply model-Ib and model-Ic to level-3 simulations, the incorporated binary orbital motions are scaled up by a factor of 3.5.

\subsubsection{model-II}

The more recent study of \cite{2017ApJS..230...15M} provides the joint distribution of $P$, $q$ and $e$ over a wide stellar mass range of main sequence stars, though in old dwarf galaxies, massive stars should have died out and the red giants used as tracers have solar mass. 

The mass ratio distribution is modeled as
\begin{equation}
\frac{\ud N}{\ud \log q} \propto q^\gamma,
\end{equation}
where
\begin{align}
  \gamma = \left\{ \begin{array}{ll}
    \gamma_\mathrm{small}, & 0.1<q<0.3;\\
    \gamma_\mathrm{large}, & 0.3<q<1.
  \end{array} \right.
 \label{eqn:ratiodist}
\end{align}

The functional form of $\gamma_\mathrm{small}$ and $\gamma_\mathrm{large}$ depends on both $m_1$ and the orbital period, $P$. The readers can refer to \cite{2017ApJS..230...15M} for details and we do not repeat the exact forms here. Note \cite{2017ApJS..230...15M} also explicitly considered the excess probability of twin binaries with mass ratios very close to unity and on the basis of Equation~\ref{eqn:ratiodist}, but when we sample the distribution of $q$, we did not consider the excess twin binary fraction. This is because at late evolutionary stages, twin binaries likely have their radii exceeding the Roche lobe size, and thus it is unlikely that there are a large fraction of twin binaries in old dwarf systems. 

The orbital period distribution of $P$ depends on both $m_1$ and $q$, which is split into two parts, $0.1<q<0.3$ and $0.3<q<1$. At $q>0.3$, the distribution depends on $m_1$, with analytical expression provided by the original paper, which we do not repeat. At $0.1<q<0.3$, the period distribution is determined by the distribution at $q>0.3$ and Equation~\ref{eqn:ratiodist}. For example, if $\frac{\ud N}{\ud \log P}=0.14$ at $q>0.3$ and $\gamma_\mathrm{small}=\gamma_\mathrm{large}=0$, we have $\frac{\ud N}{\ud \log P}=0.18$ at $q>0.1$. Here the period distribution at $P>0.1$ is the summation of the distributions at $P>0.3$ and $0.1<P<0.3$.

The distribution of the orbital eccentricity, $e$, is modeled as
\begin{equation}
\frac{\ud N}{\ud \log e} \propto e^\eta,
\end{equation}
where $\eta$ depends on both $m_1$ and $P$. We do not repeat the formula here, and the readers can check \cite{2017ApJS..230...15M} for the detailed expression.

According to \cite{2017ApJS..230...15M}, given $m_1$ and $P$, the maximum eccentricity is $e_\mathrm{max}(P)=1-(\frac{P}{2\mathrm{days}})^{-2/3}$, which guarantees the binaries to have Roche-lobe fill-factors $<\sim$70\% at periastron. 

For model-II, we sample $m_1$ from the Kroupa initial mass function only, but we adopt the binary fractions of both 70\% (model-IIa) and 36\% (model-IIb). The fraction of 36\% is in fact the integrated binary fraction for solar mass binaries by \cite{2017ApJS..230...15M}. 

For model-Ia, model-Ib, model-Ic, model-IIa and model-IIb, we do not include observational errors. On the basis of model-IIa and model-IIb, we will additionally include a typical observational error of 3~km/s to the LOSVs of mock stars. This is achieved by shifting the ``observed'' velocities (original $+$ binary motions) by an amount sampled from a Gaussian distribution with zero mean and 3~km/s of scatter. The models are then called model-IIa-err and model-IIb-err. Further on the basis of model-IIa-err and model-IIb-err, we will discard tracer stars whose changes in their LOSVs are greater than 5~km/s in two observations across one year of time. This is because stars with large changes in their LOSVs are likely binaries, and here we want to investigate the efficiency of using two epoch observations to exclude binaries given different models and observational errors. We call them model-IIa-2epoch and model-IIb-2epoch. Note for the one year of time separation between the two observations, we did not apply any scaling.

Note the binary orbital motions of model-Ib, model-Ic, model-IIa, model-IIb, model-IIa-err, model-IIb-err, model-IIa-2epoch and model-IIb-2epoch are all scaled up by a factor of 3.5. This enables us to effectively investigate intrinsic velocity dispersions of 3$<\sigma_{v_r,\mathrm{eff}}<$9~km/s. In the end and in order to investigate the range of 1$<\sigma_{v_r,\mathrm{eff}}<$3~km/s, we try models in which we scale the binary motions by a factor of 10.5. The models are included on the basis of model-IIa-2epoch and model-IIb-2epoch, and are denoted as model-IIa-FD and model-IIb-FD. Here ``FD'' stands for ``Faint enD''. Their orbital element distribution model and binary fractions are all the same as model-IIa and model-IIb, but we have excluded stars whose changes in their LOSVs are greater than 5~km/s in two observations across one year of time. Besides, for model-IIa-FD and model-IIb-FD we incorporate observational errors of 1~km/s instead of 3~km/s. This is because we are investigating 1 to 3~km/s of region with this model, and 3~km/s of error is too large, i.e., cannot enable reasonable dynamical constraints for dwarfs with intrinsic velocity dispersions between 1 and 3~km/s.  

\begin{table*}
\caption{Models adopted for sampling binary orbital motions. In the second column, we provide the references for the binary orbital element distributions, which are used to sample the changes in line-of-sight velocities (LOSVs) due to binary motions. We also summarize for each model the binary fraction ($f_\mathrm{binary}$), how the stellar mass of the primary star is sampled, whether 3-$\sigma$ clipping is applied to the ``observed'' LOSVs, the resolution of the \textsc{auriga} simulation to which the model is applied, whether observational errors are incorporated and whether we discard star particles whose change in their LOSVs are greater than 5~km/s based on two epoch ``observations'' spanning one year of time.}
\begin{center}
\begin{tabular}{lccccccc}\hline
\hline
\multicolumn{1}{c}{Model} & \multicolumn{1}{c}{reference} & \multicolumn{1}{c}{$f_\mathrm{binary}$} & \multicolumn{1}{c}{$m_1$}  & \multicolumn{1}{c}{clipping} & \multicolumn{1}{c}{resolution} & \multicolumn{1}{c}{error} & \multicolumn{1}{c}{epoch} \\ 
\hline
model-Ia & Duquennoy \& Mayor (1991) & 70\% & $0.8\msun$ & no & level-2 & no & 1 \\
model-Ib & Duquennoy \& Mayor (1991) & 70\% & Kroupa IMF & yes & level-3 & no & 1 \\
model-Ic & Duquennoy \& Mayor (1991) & 36\% & Kroupa IMF & yes & level-3 & no & 1 \\
model-IIa & \cite{2017ApJS..230...15M} & 70\% & Kroupa IMF & yes & level-3, 3.5 scaling & no & 1 \\
model-IIb & \cite{2017ApJS..230...15M} & 36\% & Kroupa IMF & yes & level-3, 3.5 scaling & no & 1 \\
model-IIa-err & \cite{2017ApJS..230...15M} & 70\% & Kroupa IMF & yes & level-3, 3.5 scaling & 3~km/s & 1 \\
model-IIb-err & \cite{2017ApJS..230...15M} & 36\% & Kroupa IMF & yes & level-3, 3.5 scaling & 3~km/s & 1 \\
model-IIa-2epoch & \cite{2017ApJS..230...15M} & 70\% & Kroupa IMF & yes & level-3, 3.5 scaling & 3~km/s & 2 \\
model-IIb-2epoch & \cite{2017ApJS..230...15M} & 36\% & Kroupa IMF & yes & level-3, 3.5 scaling & 3~km/s & 2 \\
model-IIa-FD & \cite{2017ApJS..230...15M} & 70\% & Kroupa IMF & yes & level-3, 10.5 scaling & 1~km/s & 2 \\
model-IIb-FD & \cite{2017ApJS..230...15M} & 36\% & Kroupa IMF & yes & level-3, 10.5 scaling & 1~km/s & 2 \\
\hline
\label{tbl:binary_model}
\end{tabular}
\end{center}
\end{table*}      

In both model-I and model-II, we require the stellar radius to be smaller than the Roche radius. The stellar radius is estimated from the effective temperature and the luminosity of a 10~Gyr age and solar metallicity Parsec stellar evolution isochrone, and the Roche radius is calculated from the analytical formula of \cite{1983ApJ...268..368E} at periastron, with a correction for a weak dependence on orbital eccentricity \citep{2010ApJ...721.1142M}. This is an important step, which eliminates a significant fraction of close binaries with small orbital periods and high orbital velocities. 

In Table~\ref{tbl:binary_model}, we summarize the models used to sample binary orbital motions. In particular, for model-Ib, model-Ic and all model-II, we include 3-$\sigma$ clippings to exclude stars with extreme velocities, after including the binary orbital motions. For models with observational errors, 3-$\sigma$ clippings are achieved after including errors. 
However, for initial checks with model-Ia applied to level-2 suite of simulations, we did not include such 3-$\sigma$ clippings.

\subsection{Mock dwarf images and Multi-Gaussian Expansion}
\label{sec:mockimage}

In our dynamical modeling approach (see Section~\ref{sec:methods} for details), the potential and density distributions of the luminous stellar component will be directly inferred from the optical images of the dwarfs, with the stellar-mass-to-light ratios ($M/L$) being free parameters. The image will be deprojected based on the distance and inclination angle of the dwarf. The inclination angle and the distance of the dwarf can be free parameters, but in our case we fix them to the true values. 

Thus we need to create mock images for our sample of dwarfs. We simply adopt the projected stellar mass density distribution to create the images, i.e., the read in each pixel is in unit of $\msun/\mathrm{pc}^2$ based on all bound star particles associated to the dwarf galaxy, so in our case the true value of $M/L$ is unity. 

Once the mock images are made, the luminous stellar mass distribution, $\Sigma(x',y')$, will be decomposed to a few different Gaussian components (Multi-Gaussian Expansion or MGE in short), in order to enable the analytical deprojection for any arbitrary $\Sigma(x',y')$ and to bring analytical solutions for any arbitrary matter distribution (see Section~\ref{sec:methods} and Paper I for more details). 


\section{Methodology}
\label{sec:methods}

Jeans Anisotropic Multi-Gaussian Expansion (\textsc{jam}) is a public source of code\footnote{https://github.com/lauralwatkins/cjam}. It is a powerful tool to constrain both the underlying matter distribution and the internal dynamics of tracers \citep[e.g.][]{2016MNRAS.462.4001Z,2016MNRAS.463.1117Z}, based on either LOSVs or proper motions of tracers. In this paper, we will only use the LOSVs for dynamical modeling. The version of \textsc{jam} we use is slightly different from the public version of the \textsc{jam} model for discrete data \cite{2013MNRAS.436.2598W}, with improved python interface and plotting tools. Details about \textsc{jam} can be found in \cite{2008MNRAS.390...71C} and \cite{2013MNRAS.436.2598W}, and here we only briefly introduce the method. 

The method is based on solving the axis-symmetric Jeans equation in an intrinsic frame defined on the 
dwarf galaxy with cylindrical coordinates, to solve for the first and second velocity moments.

\begin{equation}
    \frac{\nu(\overline{v_R^2}-\overline{v_\phi^2})}{R}+\frac{\partial \nu \overline{v_R^2}}{\partial R}+\frac{\partial \nu \overline{v_R v_z}}{\partial z}=-\nu \frac{\partial \Phi}{\partial R}
    \label{eqn:jeans1}
\end{equation}

\begin{equation}
    \frac{\nu \overline{v_R v_z}}{R}+\frac{\partial \nu \overline{v_R v_z}}{\partial R}+\frac{\partial \nu \overline{v_z^2}}{\partial z}=-\nu \frac{\partial \Phi}{\partial z},
    \label{eqn:jeans2}
\end{equation}
where $\nu$ is the tracer density distribution. $\Phi$ is the total potential. 
Upon solving the equation to obtain unique solutions, the cross velocity terms are assumed to be zero, i.e., $\overline{v_R v_z}=0$. In addition, the anisotropy parameter, $b$, is assumed to be constant and defined as
$\overline{v_R^2}=b\overline{v_z^2}$. A rotation parameter, $\kappa$, is introduced as $\overline{v_\phi}=\kappa
(\overline{v_\phi^2}-\overline{v_R^2})^{1/2}$. 

In our analysis, we define the $z$-axis of the intrinsic frame as the direction of the averaged spin of all bound star particles to the dwarf in the simulation, and the intrinsic frame is a right handed system. The intrinsic frame is linked to the observing frame (see Section~\ref{sec:star} above) through the inclination angle, $i$, of the dwarf galaxy

\begin{equation}
    \left( \begin{array}{c}
        x' \\
        y' \\
        z'
    \end{array} \right) = \left( \begin{array}{ccc}
        1 & 0 & 0 \\
        0 & - \cos i & \sin i \\
        0 & \sin i & \cos i
    \end{array} \right) \left( \begin{array}{c}
        x \\
        y \\
        z
    \end{array} \right),
\end{equation}
and
\begin{equation}
    \left( \begin{array}{c}
        v_{x'} \\
        v_{y'} \\
        v_{z'}
    \end{array} \right) = \left( \begin{array}{ccc}
        \cos i & - \sin i & 0 \\
        \sin i & \cos i & 0 \\
        0 & 0 & 1
    \end{array} \right) \left( \begin{array}{c}
        v_R \\
        v_\phi \\
        v_z
    \end{array} \right),
\end{equation}
where $R=\sqrt{x^2+y^2}$.

The total potential, $\Phi$, on the right hand side of Equations~\ref{eqn:jeans1} and \ref{eqn:jeans2}, is contributed by both luminous and dark matter. As we have mentioned, the luminous matter distribution is directly inferred from the surface brightness of the dwarf galaxy (see Section~\ref{sec:mockimage} above). To model the density profile of dark matter, we adopt in our analysis a double power law functional form of

\begin{equation}
    \rho(r)=\frac{\rho_s}{(r/r_s)^\gamma (1 +r/r_s)^\alpha},
    \label{eqn:double}
\end{equation}
with the model parameters ($\rho_s$, $r_s$ and $\gamma$) to be constrained. Note in our analysis throughout this paper, the outer power law index, $\alpha$, will be fixed to 3. 

In order to have analytical solutions for any given potential model and tracer distribution, MGE is not only applied to the 2-dimensional surface density distribution of the luminous stellar component (see Section~\ref{sec:star} above), but also to the underlying model for the dark matter distribution and to the density distribution of tracers ($\nu$) as well\footnote{In our case, tracers and the luminous stellar component have the same distribution, and therefore the same MGEs. Note the normalization of the MGE components for tracers is not important, which cancels out on two sides of the equations.}. Each MGE component would have analytical solutions to Equations~\ref{eqn:jeans1} and \ref{eqn:jeans2}. In principle, each MGE component of the tracer population can have its own rotation parameter, $\kappa_k$, and velocity anisotropy parameter, $b_k$. $M/L$ for each MGE component can also differ, but in our analysis we treat $\kappa$, $b$ and $M/L$ to be the same for different MGEs. 

For an observed star with position $\boldsymbol{x'}_i=(x'_i,y'_i)$ on the image plane, which has observed velocity 
$\boldsymbol{v}_i=\left(v_{x',i},v_{y',i},v_{z',i}\right)$ and error matrix of 

\begin{equation}
  \boldsymbol{S}_i =
  \left( {\begin{array}{ccc}
    \sigma^2_{v_{x'},i} & 0 & 0 \\
    0 & \sigma^2_{v_{y'},i} & 0 \\
    0 & 0 & \sigma^2_{v_{z'},i} \\
  \end{array} } \right),
\end{equation}
its position, $\boldsymbol{x'}_i$, can be transformed to the intrinsic frame to solve the corresponding velocities 
and velocity dispersions, based on a set of model parameters, $\Theta$. Solution for each MGE is sought, and solutions 
of different MGEs are added together in the end. The solutions are then transformed back to the observing frame. The 
mean velocity predicted by the model in the observing frame is denoted as $\boldsymbol{\mu}_i=\left( v_{x',i},
v_{y',i},v_{z',i} \right)$, and the covariance matrix is defined through both the first and the second velocity moments

\begin{align}
    &\boldsymbol{C}_i = \nonumber\\
    &\left( \begin{array}{ccc}
        \vsqxm - \vxm^2 & \vsqxym - \vxm\,\vym & \vsqxzm - \vxm\,\vzm \\
        \vsqxym - \vxm\,\vym & \vsqym - \vym^2 & \vsqyzm - \vym\,\vzm \\
        \vsqxzm - \vxm\,\vzm & \vsqyzm - \vym\,\vzm & \vsqzm - \vzm^2
    \end{array} \right). 
\end{align}

\begin{figure*} 
\begin{center}
\includegraphics[width=0.8\textwidth]{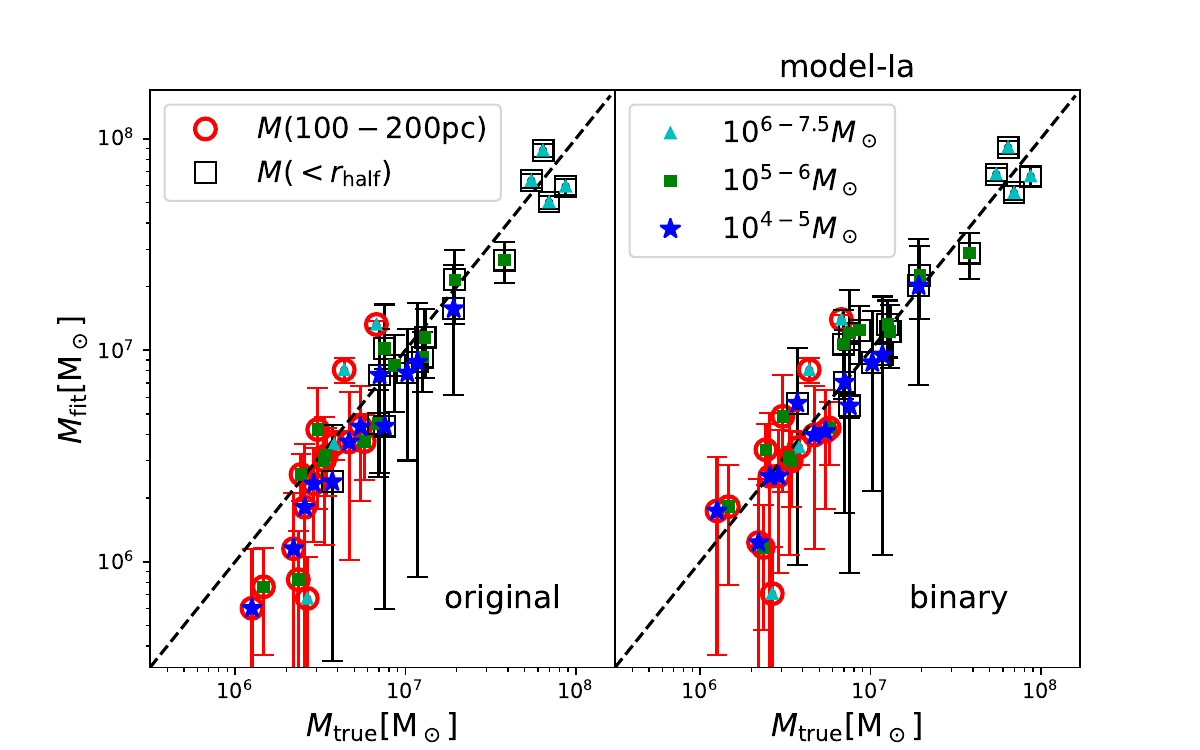}%
\end{center}
\caption{Best-fitting versus true mass, for the total masses enclosed within the half-mass radius of tracers (black square, $M(<r_\mathrm{half})$) and the masses between 100 to 200~pc (red circles, $M(100-200\mathrm{pc})$). This is shown for 17 dwarf systems selected from the \textsc{auriga} level-2 simulations. Results in the left panel are dynamical constraints based on the true velocities of tracer star particles in the simulation, while results in the right panel are based on the velocities after incorporating binary orbital motions (model-Ia). In both panels, the diagonal black dashed line marks $y=x$ to guide the eye. Error bars are 1-$\sigma$ statistical errors of the best fits. In the left panel, most measurements are ensemble unbiased, which distribute symmetrically around the black dashed line. However, at the low-mass end of the left panel, the best-fitting masses are significantly under-estimated, which could be either due to the small tracer sample size or due to the softening scale limit. In the right panel, red circles or black squares over plotted with cyan triangles, green squares and blue stars are dwarf galaxies with different total stellar masses, as indicated by the legend.}
\label{fig:level2overall}
\end{figure*}

\begin{figure*}
\begin{center}
\includegraphics[width=0.8\textwidth]{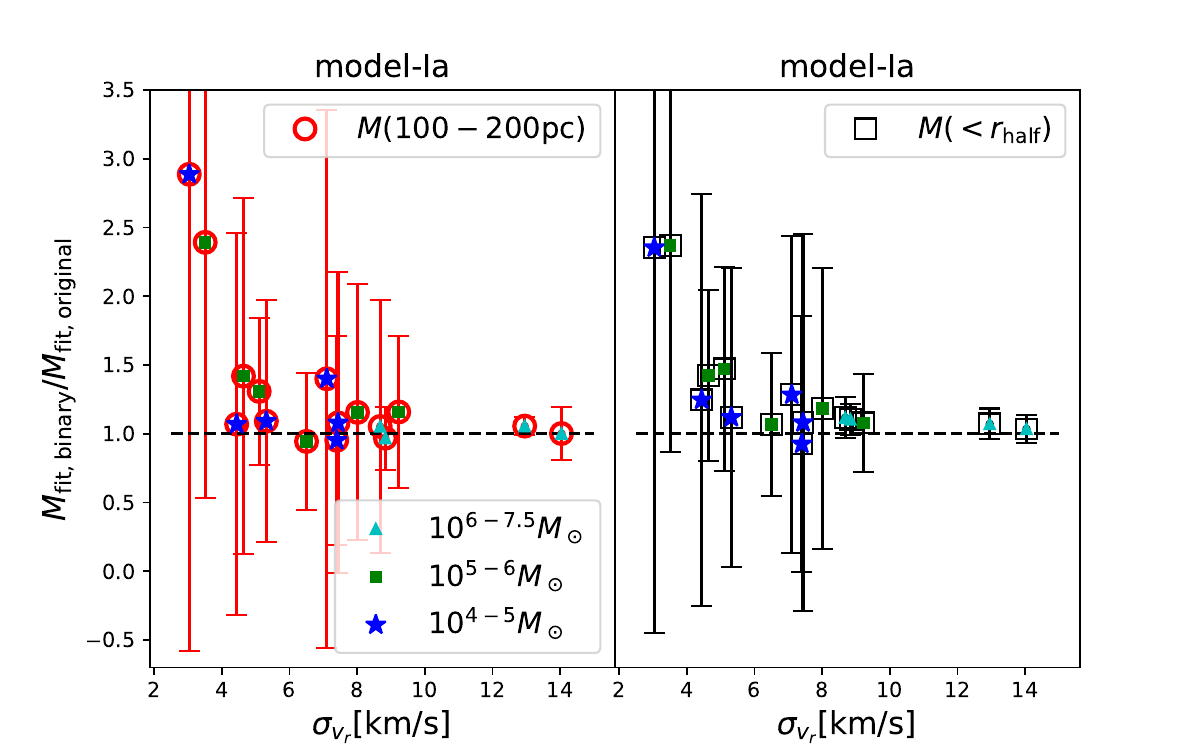}%
\end{center}
\caption{The ratios between best-fitting total masses after and before including binary motions of model-Ia ($y$-axis), reported as a function of the intrinsic line-of-sight velocity dispersions of the dwarf systems ($x$-axis). We show the total masses enclosed within the half-mass radius of tracers (black square, left panel, $M(<r_\mathrm{half})$) and the masses between 100 to 200~pc (red circles, right panel, $M(100-200\mathrm{pc})$). This figure is based on 17 dwarf systems from the \textsc{auriga} level-2 simulations. Red circles or black triangles with cyan triangles, green squares and blue stars are dwarf galaxies with different total stellar masses, as indicated by the legend. Error bars are 1-$\sigma$ statistical errors of the best fits.}
\label{fig:level2ratio}
\end{figure*}

By assuming the velocity distribution predicted by the model is a tri-variate Gaussian with mean velocity 
$\boldsymbol{\mu}_i$ and covariance $\boldsymbol{C}_i$ at $\boldsymbol{x'}_i$, the likelihood can be written 
as

\begin{align}
    L_i^\mathrm{dwarf} & = p \left( \boldsymbol{v}_i | \boldsymbol{x'}_i, \boldsymbol{S}_i, \Theta
        \right) \nonumber \\
        & = p \left( \boldsymbol{v}_i | \boldsymbol{x'}_i, \boldsymbol{S}_i, \boldsymbol{\mu}_i, \boldsymbol{C}_i
        \right) \nonumber \\
    & = \frac{ \exp \left[ - \frac{1}{2} \left( \boldsymbol{v}_i - \boldsymbol{\mu}_i
        \right)^{\mathrm{T}} \left( \boldsymbol{C}_i + \boldsymbol{S}_i \right)^{-1} \left( \boldsymbol{v}_i -
        \boldsymbol{\mu}_i \right) \right] } { \sqrt{ \left( 2 \pi \right)^{3} \left|
        \left( \boldsymbol{C}_i + \boldsymbol{S}_i \right) \right| } }.
    \label{eqn:like}
\end{align}

The total likelihood is the product of the likelihood for each star

\begin{equation}
    L=\prod_{i=1}^{N_\mathrm{star}}L_i.
\end{equation}

Note, however, in our case, we only use the LOSV information, while the information of $v_{x',i}$ and $v_{y',i}$ are not available. We simply set $v_{x',i}=v_{y',i}=0$ and input very large values for $\sigma^2_{v_{y'},i}$ and  $\sigma^2_{v_{z'},i}$. This is equivalent to only fit the observed first and second moments of LOSVs.

The list of parameters used in our modeling are summarized in the following:

(1) Rotation parameter, $\kappa$;

(2) Velocity anisotropy, $b$;

(3) Dark matter halo scale density, $\rho_s$;

(4) Dark matter halo scale radius, $r_s$;

(5) Inner density slope of the host dark matter halo, $\gamma$;

In Paper I, we have reported strong degeneracies between the stellar and dark matter components, and thus $M/L$ is poorly constrained for our sample of dwarf galaxies. Observationally, $M/L$ can be alternatively constrained through stellar population synthesis modeling and then fixed upon dynamical modeling. Hence in our analysis through this paper, the stellar-mass-to-light ratio, $M/L$, will simply be fixed to unity, i.e, its true value. We also fix the distance and the inclination angle to be their true values. Moreover, the outer density slope, $\alpha$, will be fixed to $3$, but we have also tried to vary the outer slopes, and our conclusions are not sensitive to $\alpha$.

\section{Results}
\label{sec:results}

\subsection{model-Ia applied to level-2 resolution}

\begin{figure}
\begin{center}
\includegraphics[width=0.49\textwidth]{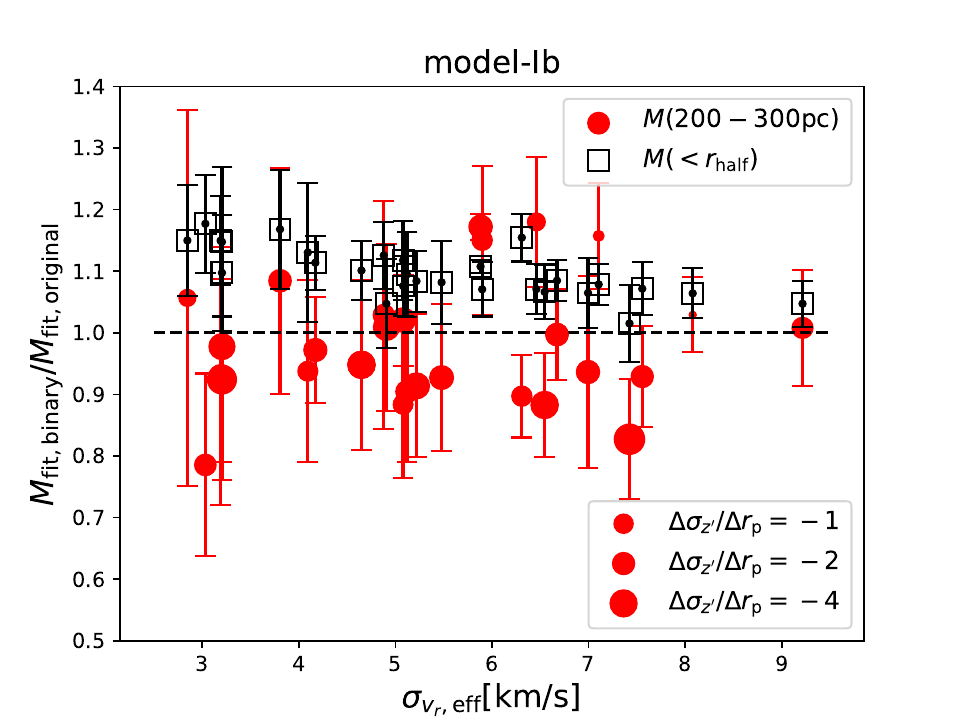}%
\end{center}
\caption{The ratios between best-fitting total masses after and before including binary motions of model-Ib ($y$-axis), reported as a function of the intrinsic effective line-of-sight velocity dispersions of the dwarf systems ($x$-axis). We show the total masses enclosed within the half-mass radius of tracers (black square, $M(<r_\mathrm{half})$) and the masses between 200 to 300~pc (red dots, $M(200-300\mathrm{pc})$). The assumed $f_\mathrm{binary}$ is 70\%. This figure is based on 28 dwarf systems from the \textsc{auriga} level-3 simulations, after scaling the line-of-sight velocities by a factor of 3.5 to represent low-mass dwarf galaxies (see Section~\ref{sec:dwarf} for details). Note the radial scales (200 and 300~pc) are based on original coordinates from the simulation, which are not scaled. The size of red dots is inversely proportional to the averaged radial gradient of the intrinsic line-of-sight velocity dispersion profile over 0.2$r_\mathrm{half}$ and 2$r_\mathrm{half}$ (binaries not included). The more negative the gradients are, the larger the symbol sizes. Dwarfs with $M(200-300\mathrm{pc})$ deflated are more likely to have more negative radial gradients in their velocity dispersion profiles. Error bars are 1-$\sigma$ statistical errors of the best fits.}
\label{fig:level3scale70}
\end{figure}

We first show results based on 17 dwarf galaxies selected from the level-2 suite of \textsc{auriga} simulations. The results in this current subsection are based on model-Ia. Figure~\ref{fig:level2overall} shows the best constrained masses by \textsc{jam} ($y$-axis) versus the true masses ($x$-axis). In the two panels, the true masses along the $x$-axis are exactly the same, whereas the \textsc{jam} constrained masses are based on the original LOSVs of star particles and the LOSVs after including binaries, for the left and right panels, respectively. Red circles represent the masses between 100 and 200~pc, $M(100-200\mathrm{pc})$, and black squares refer to the masses within the half-mass radius of tracer stars, $M(<r_\mathrm{half})$. Note the softening scale of \textsc{auriga} level-2 resolution is only slightly smaller than 100~pc, and the deviation from the Newtonian gravitational potential starts to become important within the quoted softening scale \citep{2010MNRAS.401..791S}.

The $y=x$ diagonal dashed lines are to guide the eye, and are exactly the same in both panels. For massive dwarfs, there is no prominent difference between the best-constrained mass in the two panels. However, for the few red circles and black squares at the low mass end, their \textsc{jam} constrained masses along the $y$-axis are prominently higher after including binary orbital motions in the right panel than the left one. This indicates such low-mass systems are more likely to be significantly affected by binaries upon dynamical constraints. 

In the right panel, we further mark the total stellar mass of different dwarf systems. Symbols over plotted with cyan triangles, green squares and blue stars represent dwarf systems with total stellar mass in the ranges of $10^{6-7.5}\msun$, $10^{5-6}\msun$ and $10^{4-5}\msun$, respectively. Despite the fact that the blue stars have the smallest total mass, they span a wide range in $M_\mathrm{fit}(100-200\mathrm{pc})$ and $M_\mathrm{fit}(<r_\mathrm{half})$. It is not really the total stellar mass matters, but instead, the amount of inflations for the mass within a given radius is more closely related to the velocity dispersion at the corresponding radius. 

In both panels, the symbols distribute more or less symmetrically around the $y=x$ diagonal line, indicating the best-fitting dynamical mass is ensemble unbiased. However, at the low mass end of the left panel, the few red circles are prominently lower than the diagonal line, indicating systematic under-estimates in the dynamical mass than the truth. After including binaries, the few red circles at the low-mass end go more symmetrically around the diagonal line. We think the under-estimates in the left panel could be related to the softening scale of the level-2 resolution. Although $100-200$~pc is greater than the softening scale of level-2, the half-light radius of such low-mass dwarfs is a few hundred pc \citep{2021MNRAS.507.4953G}, with an average of $\sim$~300~pc at $10^{4-5}\msun$. The softening scale is significant compared with the dwarf size, and thus internal dynamics of star particles for such low-mass dwarf galaxies is likely still affected above the softening length. Moreover, such low-mass level-2 dwarf systems have only a few tens of star particles, which could be more vulnerable to the resolution limit. The inclusion of binary motions, on the other hand, boost the best constrained dynamical mass. The boosted dynamical masses better go through the $y=x$ diagonal line at the low-mass end, which could just be a coincidence due to the co-addition of two effects. 

Despite the possible effect by the resolution limit, the difference between the left and right panels at the low-mass end due to binaries is real and robust. Figure~\ref{fig:level2ratio} further shows the ratios between the \textsc{jam} constrained dynamical masses, after and before the inclusion of binaries. The ratios are reported as a function of the intrinsic LOSV dispersions of the dwarfs (without binary). Left and right panels show $M(100-200\mathrm{pc})$ and $M(<r_\mathrm{half})$, respectively. There exists a prominent trend that with the decrease in $\sigma_{v_r}$, the binary motions tend to introduce larger inflations in the best-constrained dynamical masses. At $\sigma_{v_r}\sim 3$~km/s, the amount of boost in $M(100-200\mathrm{pc})$ or $M(<r_\mathrm{half})$ based on model-Ia can be a factor of 2.5. Note, however, this factor of 2.5 is without 3-$\sigma$ clippings of the observed LOSVs. After 3-$\sigma$ clippings, it gets much smaller. Unfortunately, we have only a few tens of star particles as tracers here, so the associated error bars are very large. 

To summarize, Figures~\ref{fig:level2overall} and \ref{fig:level2ratio} unambiguously show us the effect of how binary orbital motions boost the best-constrained dynamical mass of dwarf galaxies at low $\sigma_{v_r}$. With a few tens of star particles as tracers, the trends are prominent at the low-mass end. However, such a small number of tracers lead to large statistical errors, and the dynamical mass seems to be affected by the resolution limit for such low-mass systems. These make the results hard to interpret. Thus in the following subsections, we move on to show results based on larger samples of tracer star particles from more massive dwarf galaxies of the \textsc{auriga} level-3 simulations, with a scaling method to manually increase the level of binary motions. 


\subsection{model-Ib and model-Ic applied to scaled level-3 resolution}

\begin{figure*} 
\begin{center}
\includegraphics[width=0.9\textwidth]{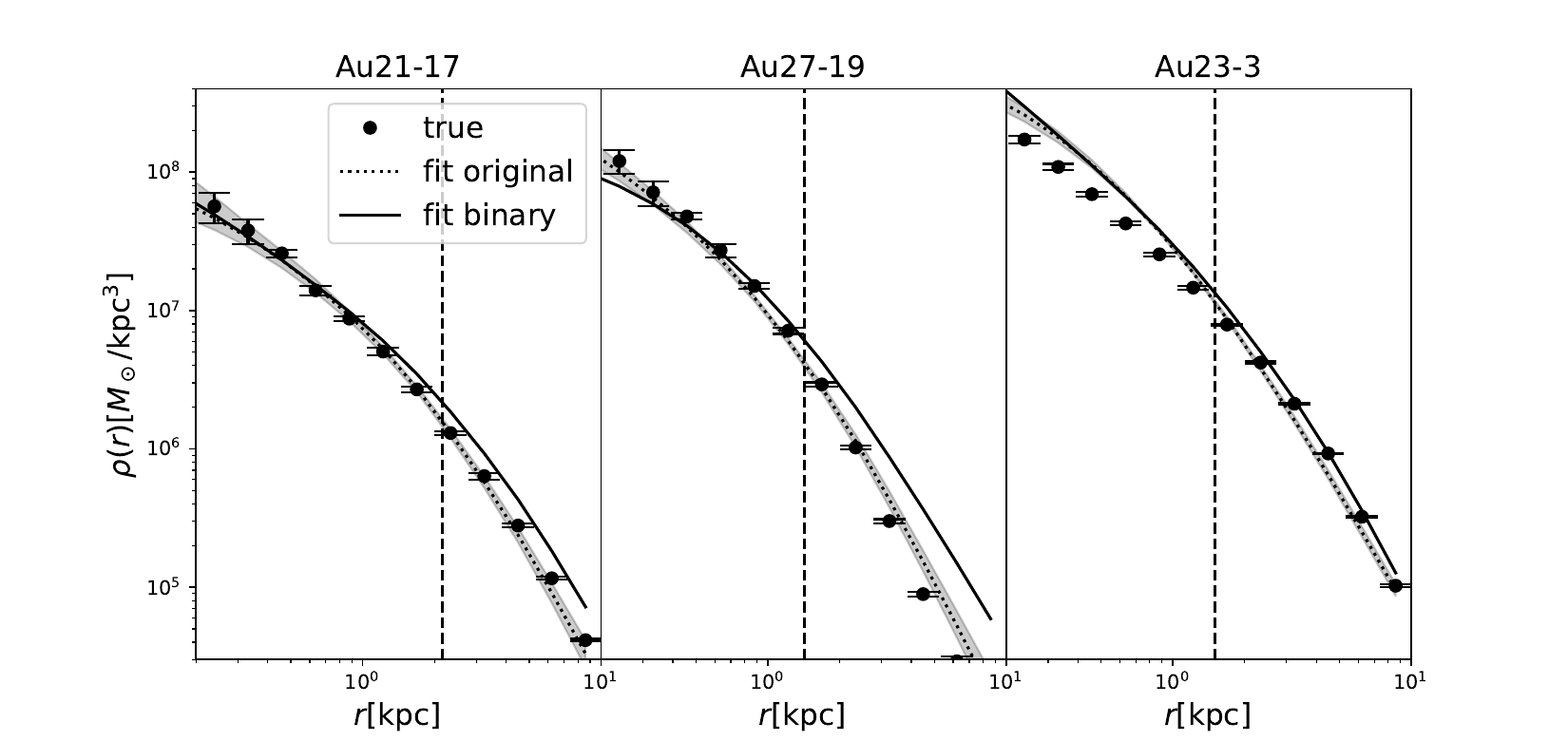}%
\end{center}
\caption{Black dots with error bars are true total density profiles in three example dwarf systems from the level-3 resolution of \textsc{auriga}, Au21-17 (left) and Au27-19 (middle) and Au23-3 (right). The error bars are the 1-$\sigma$ scatters of 100 bootstrapped samples. Black dotted lines are \textsc{jam} constrained density profiles from stellar dynamics, with the black shaded region reflecting the 1-$\sigma$ uncertainty region. Black solid lines are \textsc{jam} constrained density profiles after including binary orbital motions from model-Ib. The errors of the black solid lines are similar to that of black dotted lines, and are thus not shown. In the right plot, the inflation in $M(200-300\mathrm{pc})$ is greater than the inflation in $M(<r_\mathrm{half})$, which is likely due to the few number of tracer star particles in the center and statistical fluctuations. Note the $x$-axis radius are based on original coordinates from the simulation, which are not scaled. In all three panels, the vertical dashed line indicates the position of $r_{\rm half}$.}
\label{fig:exampleprof}
\end{figure*}

\begin{figure*} 
\begin{center}
\includegraphics[width=0.9\textwidth]{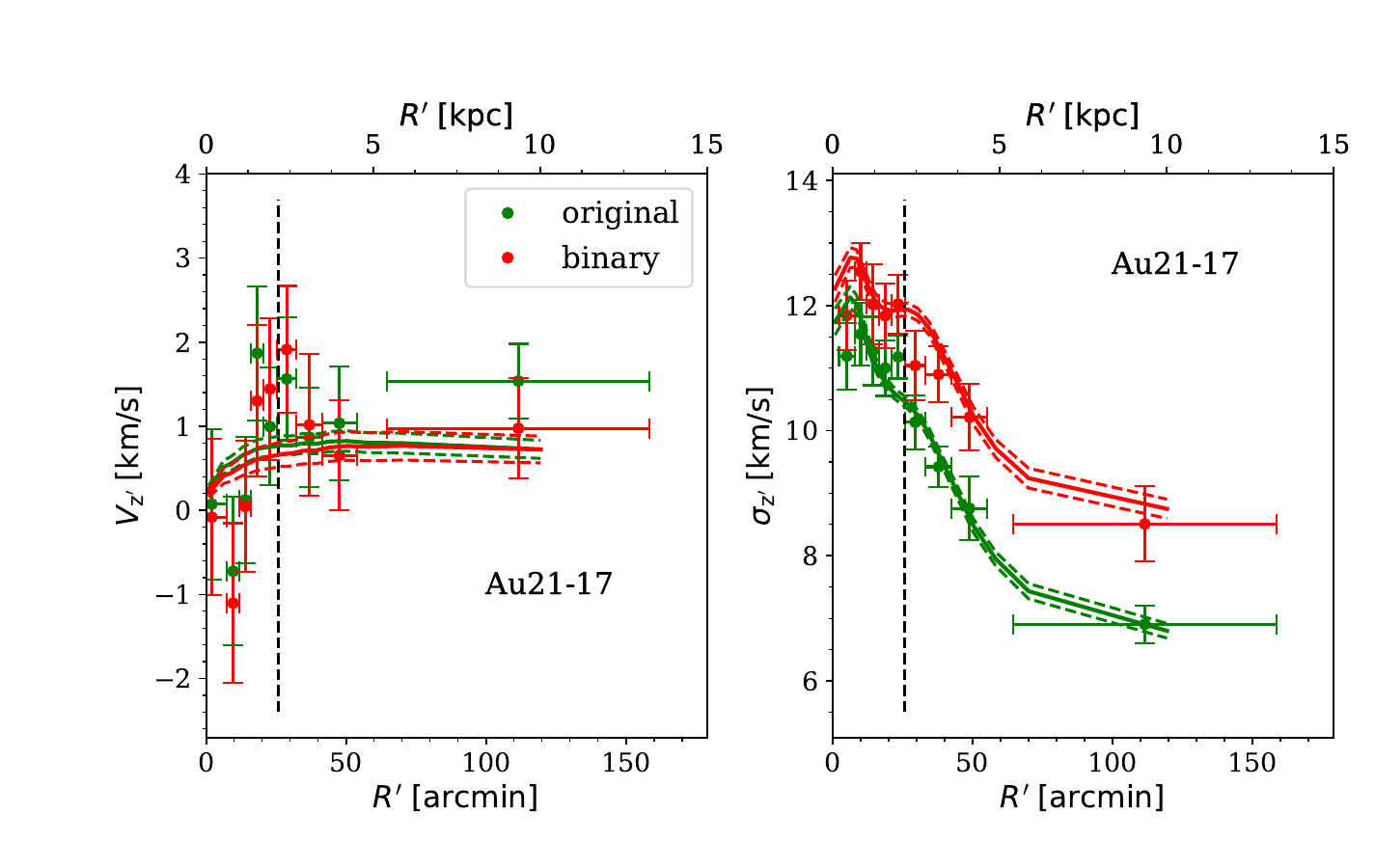}\\
\includegraphics[width=0.9\textwidth]{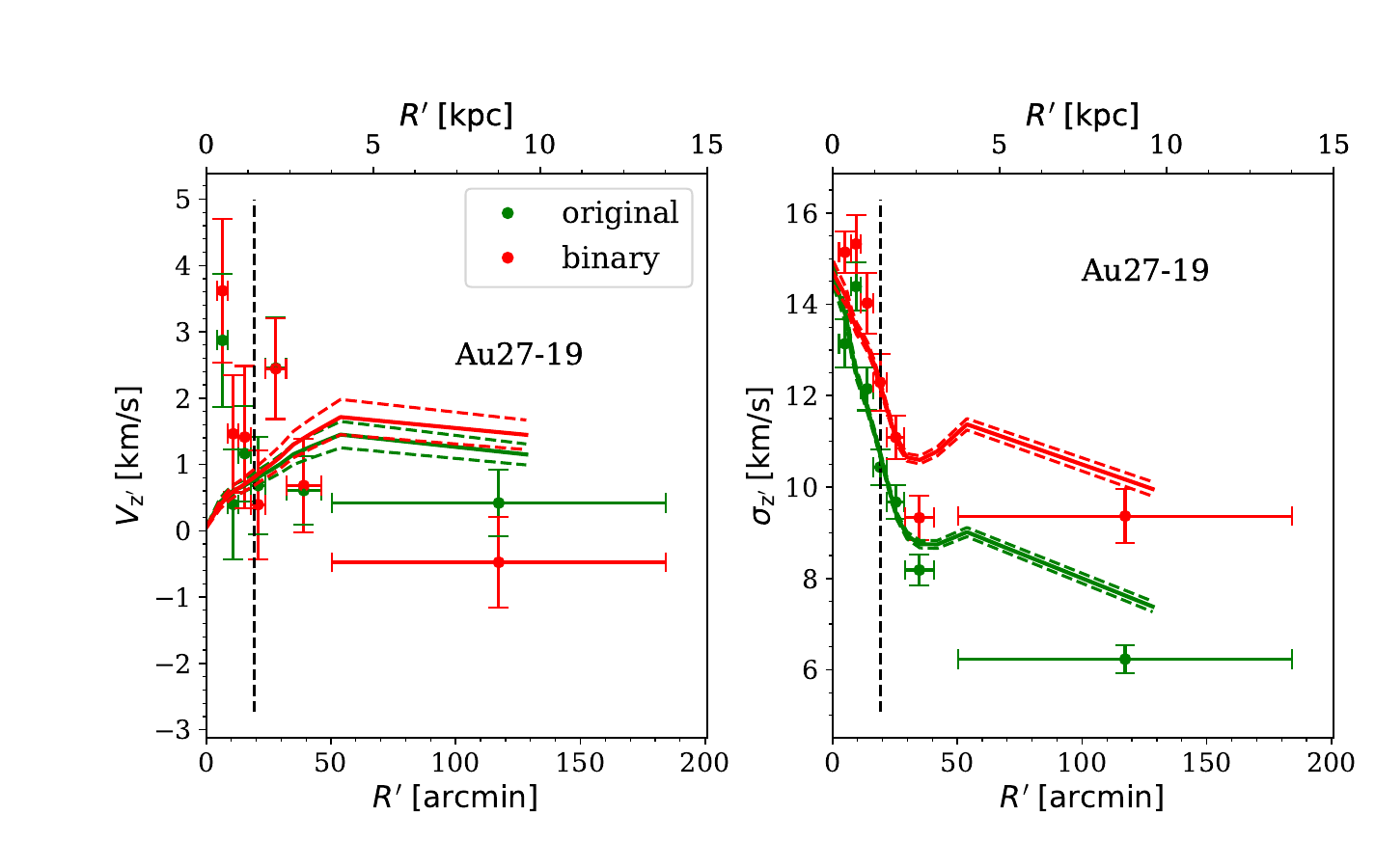}%
\end{center}
\caption{The first (left panel) and second (right panels) moments of the line-of-sight velocities along the major axes of Au21-17 (top) and Au27-19 (bottom). In all plots and panels, green and red dots are velocity moments based on star particles within sectors of $\pm45$ degrees to the major axes. Each bin contains 200 stars. Binary orbital motions are not included for green dots, and are included for red dots. Solid curves with corresponding colors are best fits by \textsc{jam}, with dashed curves around the solid ones showing the 1-$\sigma$ uncertainties of the best-fitting models. Note in both plots, the velocity dispersions are the intrinsic values from the simulations and are not scaled. The vertical black dashed lines mark the positions of $r_\mathrm{half}$.}
\label{fig:vgraph}
\end{figure*}

\begin{figure} 
\begin{center}
\includegraphics[width=0.49\textwidth]{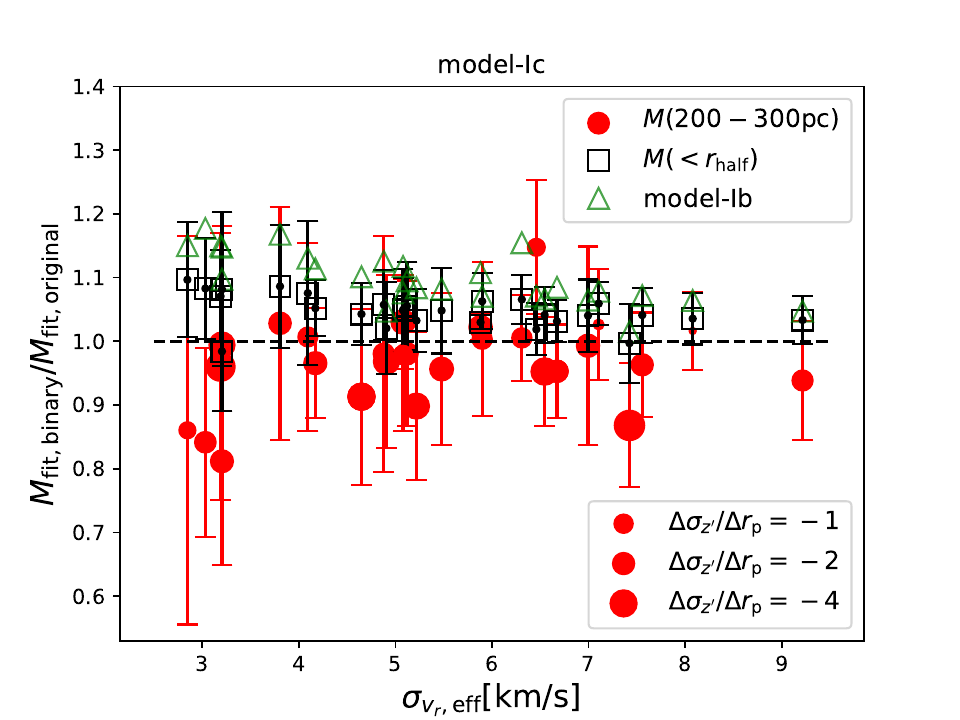}%
\end{center}
\caption{Similar to Figure~\ref{fig:level3scale70}, but is based on an assumed binary fraction of 36\% (model-Ic). Green triangles are repeats of the inflations in $M(<r_\mathrm{half})$ from model-Ib with binary fraction of 70\%, i.e., exactly the same as the black squares in Figure~\ref{fig:level3scale70}.}
\label{fig:level3scale36}
\end{figure}

\begin{figure*} 
\begin{center}
\includegraphics[width=0.49\textwidth]{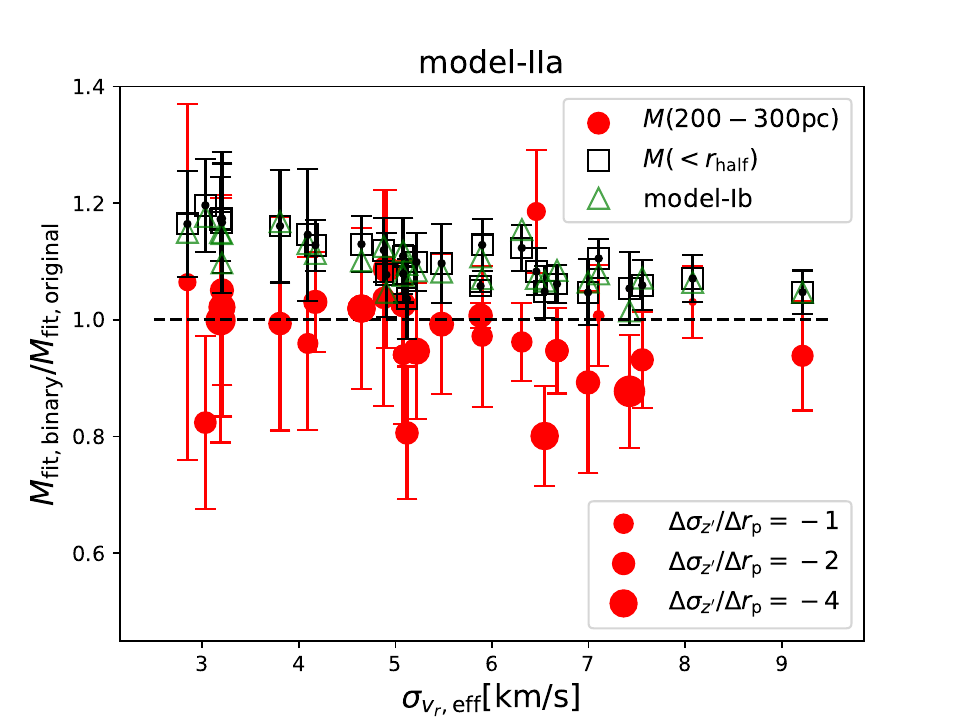}%
\includegraphics[width=0.49\textwidth]{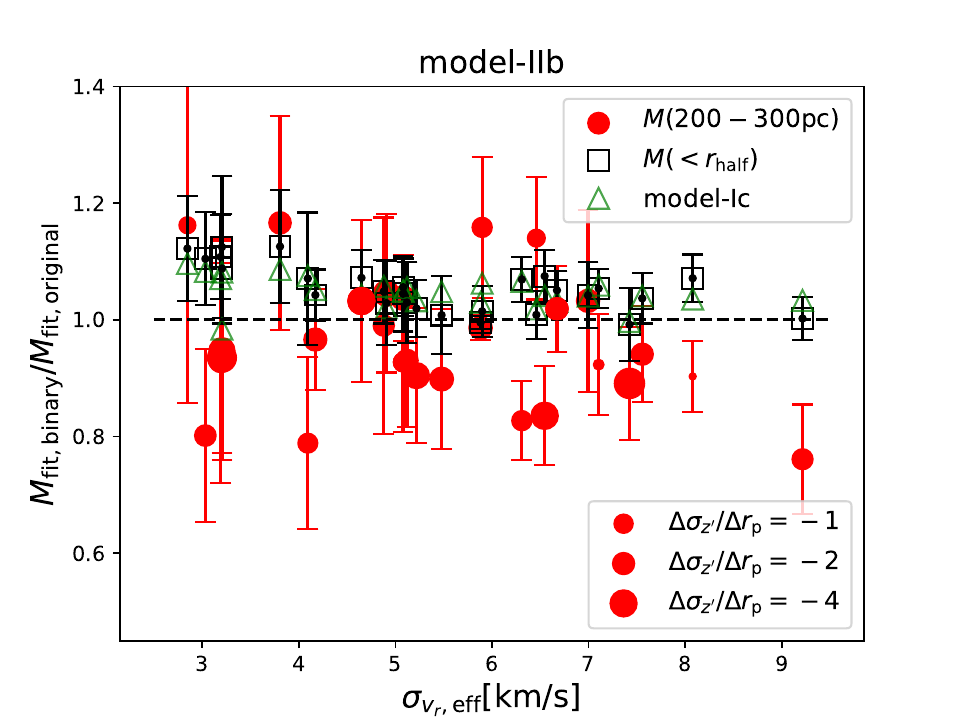}%
\end{center}
\caption{{\bf Left:} Similar to Figure~\ref{fig:level3scale70}, but is based on model-IIa. The adopted binary fraction is 70\%. {\bf Right:} Similar to Figure~\ref{fig:level3scale36}, but is based on model-IIb. The adopted binary fraction is 36\%. Green triangles in either plots show the inflations in $M(<r_\mathrm{half})$ from model-Ib (Figure~\ref{fig:level3scale70}) and model-Ic (Figure~\ref{fig:level3scale36}), with the same binary fractions with the corresponding black squares, but incorporating different binary orbital models. In both plots, the size of red dots is inversely proportional to the averaged radial gradient of the intrinsic line-of-sight velocity dispersion profile over 0.2$r_\mathrm{half}$ and 2$r_\mathrm{half}$. The more negative the gradients are, the larger the symbol sizes.}
\label{fig:level3scalemoe}
\end{figure*}

As we have discussed in Section~\ref{sec:binary_motion}, the \textsc{auriga} level-3 resolution is not enough to bring enough number of tracer star particles for decent dynamical modelings of $10^{4-5}\msun$ dwarf systems. Thus in this subsection we combine level-3 with a scaling method. Though the resolution of level-3 is even lower, we manually increase the binary orbital motions by a factor of 3.5, while incorporating binary motions to level-3 dwarf systems more massive than $10^{7.5}\msun$. This is equivalently to having decreased $\sigma_{v_r}$ of level-3 massive dwarf systems by a factor of 3.5, and thus we can investigate the effect of binary motions on dwarf systems with effective line-of-sight velocity dispersions, $\sigma_{v_r,\mathrm{eff}}$, in the range of 3 to 9~km/s (see the distributions in Figure~\ref{fig:vhist}), and at the same time we have enough number of star particles to be used as tracers. 

After applying model-Ib to 28 such dwarf systems from level-3, the results are shown in Figure~\ref{fig:level3scale70}. Similar to Figure~\ref{fig:level2ratio}, the $y$-axis demonstrates the ratios between the \textsc{jam} constrained dynamical masses after and before including binary motions, which are reported as a function of the intrinsic $\sigma_{v_r,\mathrm{eff}}$ of the host dwarf systems. Red dots and black squares refer to the total masses between 200 and 300~pc, $M(200-300\mathrm{pc})$, and $M(<r_\mathrm{half})$, respectively, but now they are shown in the same panel. Here the softening scale of level-3 resolution is close to 200~pc at $z=0$, so instead of plotting $M(100-200\mathrm{pc})$, we adopt $M(200-300\mathrm{pc})$. 

Here we only show the ratios between the best constrained masses before and after including binaries. The readers may wonder what is the performance of \textsc{jam} without including binaries, as compared to the truth in the simulation, which we refer the readers to our Paper I for details, in which we performed very detailed comparisons of the best-fiting mass profiles by \text{jam} and the truth. In brevity, we find ensemble unbiased best fits with respect to the truth, with a scatter of 0.167~dex in $M(200-300\mathrm{pc})$ for the 28 systems from the level-3 resolution.

In order to control the size of statistical errors, we use all available star particles in these dwarf systems, if the total number of bound star particles is smaller than 20,000. If the total number of star particles is greater than 20,000, we randomly draw a subsample of 20,000, in order to control the time cost in a reasonable range. 

Now with the statistical errors controlled smaller, we can clearly see the black squares deviating from unity. With the decrease in the intrinsic $\sigma_{v_r,\mathrm{eff}}$, the dynamically constrained $M(<r_\mathrm{half})$ is more significantly inflated, reaching maximums of $\sim15$\% at $\sigma_{v_r,\mathrm{eff}}\sim3$~km/s. With $f_\mathrm{binary}$ as high as 70\%, the dynamically constrained $M(<r_\mathrm{half})$ can be inflated by $\sim$4\% even at velocity dispersion of 9~km/s, though the lower error bar touches $\sim$1\%. Between 5 and 8~km/s, the amount of inflations are $\sim$6-11\%. Note from now on 3-$\sigma$ clippings have been applied to the observed LOSVs, so at the low $\sigma_{v_r,\mathrm{eff}}$ end, the amount of inflations is significantly smaller than those in Figure~\ref{fig:level2ratio}. 

Compared with the black squares, the behavior of red dots is quite different in Figure~\ref{fig:level3scale70}. Surprisingly, when all black squares are above unity, we can see a large fraction of red dots are in fact below unity. In other words, instead of showing inflations after including binary motions, $M(200-300\mathrm{pc})$ are more likely {\it deflated}. Moreover, although $M(<r_\mathrm{half})$ are inflated more with the decrease in $\sigma_{v_r,\mathrm{eff}}$, the amount of deflations in $M(200-300\mathrm{pc})$ does not show prominent dependence on $\sigma_{v_r,\mathrm{eff}}$. For dwarf galaxies with $\sigma_{v_r,\mathrm{eff}}$ of 5-8~km/s and if their $M(200-300\mathrm{pc})$ are deflated due to binary motions, the amount of deflations is up to $\sim$10-20\%. 

In order to investigate the reasons why $M(200-300\mathrm{pc})$ are more likely deflated after including binary motions, we show in the left and middle plots of Figure~\ref{fig:exampleprof} the dynamically constrained density profiles before and after including binary motions, for two example dwarf systems, Au21-17 and Au27-19. The two systems correspond to the symbols with the smallest $\sigma_{v_r,\mathrm{eff}}$ in Figure~\ref{fig:level3scale70} (the two left most red dots). Here Au21-17 has its $M(200-300\mathrm{pc})$ inflated (red circle above unity though with large errors), and Au27-19 has its $M(200-300\mathrm{pc})$ deflated (red circle below unity) after including binary motions. Note although we show the example density profiles for two dwarf systems only, the same trend holds for other dwarf galaxy systems from level-3.

In the left plot of Figure~\ref{fig:exampleprof}, though the solid curve is above the dotted curve at all radii, the difference is greater at larger radii, which gradually becomes smaller in inner regions. In the middle plot, the solid curve is lower in amplitude than the dotted curve within $\sim$350~pc, but at larger radii, the solid curve is more above the dotted one. It seems, the amount of inflations is more significant in outskirts, whereas it becomes much smaller in inner regions, which even turns into deflations at the very center. 

The above trend is due to the negative radial gradients in the velocity dispersion profiles of realistic dwarf galaxy systems. In Figure~\ref{fig:vgraph}, we show the LOSV and LOSV dispersion profiles for Au21-17 and Au27-19. After including binary motions, the first moments do not show any significant systematic differences, and the second moments are boosted. However, the LOSV dispersions significantly decrease with the increase in radii. As a result, the fractional inflations in the LOSV dispersion due to binary motions are more prominent in outskirts. 

Now we clearly see that the velocity dispersions show strong radial dependences. This would also cause radius dependent biases due to binary motions in dynamical constraints. The amount of intrinsic dispersions is significantly lower at larger radius, and thus $\sigma_{v_r}$ and the best constrained dynamical masses are likely to be inflated more. The larger inflations in $\sigma_{v_r}$ lead to higher density profiles in outskirts, whereas the smaller inflations in $\sigma_{v_r}$ and in central regions do not lead to as significant increases in the best-fitting density profiles. This explains why the black solid curves are more above the black dotted curves in outskirts of Figure~\ref{fig:exampleprof}, whereas the difference is much smaller in central regions, with even switched trends. Perhaps due to a reconcile to better fit the outskirts, the central density profiles can be even under estimated after including binary motions. As a result, we see many red dots are below unity in Figure~\ref{fig:level3scale70}. Moreover, since dwarf systems investigated in this study all have such negative radial gradients in their velocity dispersion profiles, this perhaps answers why the deflations in $M(200-300\mathrm{pc})$ do not show strong dependences on $\sigma_{v_r}$. 

The size of red dots in Figure~\ref{fig:level3scale70} is chosen to be inversely proportional to the averaged radial gradient of the intrinsic\footnote{Using the gradients of the LOSV dispersion profiles after including binaries lead to very similar conclusions.} LOSV dispersion profile over 0.2$r_\mathrm{half}$ and 2$r_\mathrm{half}$. Explicitly, the more negative the gradients in the LOSV dispersion profiles are, the larger the symbol sizes. Note when calculating the gradients in the LOSV dispersion profiles, we draw circles in the plane perpendicular to the LOS direction, instead of distinguishing the major and minor axes as in Figure~\ref{fig:vgraph}. We can see that for dwarfs with $M(200-300\mathrm{pc})$ deflated, they are indeed more likely to have more negative radial gradients in their LOSV dispersion profiles. This supports our explanation above. 

However, the readers may have the concern that our choice of radial range ($200-300\mathrm{pc}$) is close to the softening scale in \textsc{auriga} level-3 (slightly below 200~pc), which is marginal. Is it possible that the deflations are affected by the softening? We thus try a different choice of radial range ($300-400\mathrm{pc}$), and find the correlations between the gradients in the LOSV dispersion profiles and the amounts of inflations/deflations in $M(300-400\mathrm{pc})$ still exist. This supports our argument that the deflations are due to the shape of the velocity dispersion profiles, rather than due to numerical effects. Moreover, if there may exist any deviation from Newtonian on small scales, the effect is very likely the same before and after including binaries, which is not expected to violate our conclusion. 

There are a few red dots which are above the black squares in Figure~\ref{fig:level3scale70}. These are very likely statistical fluctuations due to the small number of tracer star particles in the very center of these systems, because these red dots still marginally agree with the black squares within 1-$\sigma$ errors. One example density profile is shown in the right plot of Figure~\ref{fig:exampleprof} (Au23-3). We can see the black solid line (best fit after including binaries) goes more above the dotted line at large radius, and the black and dotted lines gradually becomes more similar at smaller radius. In the very center, the black line starts to turn up again, but the errors also go significantly larger. 

Note similar trend was not seen for red circles in Figure~\ref{fig:level2ratio}. This is mainly because at the low $\sigma_{v_r,\mathrm{eff}}$ end, the number of tracer star particles is only a few tens in the level-2 resolution, which is not enough to resolve the radial gradient. Besides, we did not include 3-$\sigma$ clippings to the observed LOSVs in Figure~\ref{fig:level2ratio}, so the amount of inflations can be much larger, as biased by some extreme velocities near the LOSV distribution tails. 

Going back to Figure~\ref{fig:vgraph}, in the bottom right plot, we can see the velocity dispersions are not very well fit at large radii. The model is significantly higher than the actual dispersions along the major axis and beyond 30 arcmin. In the left panel, the model tends to fit a rotation of this dwarf system. This gives a good fit in central regions, but beyond 30 arcmin, the actual LOSVs do not show an as strong trend of rotation as those particles in central regions. In fact, this dwarf system is undergoing some  rotations in central regions, but no such rotations in outskirts. However, the model fails to capture such a feature, because we fix the rotation parameter, $\kappa$, to be the same for different MGE components (see Section~\ref{sec:mockimage} for details). As a result, the first and second moments are not very well fit in outskirts. Allowing $\kappa$ and $b$ to differ for different MGE components can potentially improve the fitting, but our conclusions about how binary motions inflate or deflate the dynamical constraints are not affected. 

We emphasize that $f_\mathrm{binary}$ adopted in model-Ib and Figure~\ref{fig:level3scale70} is as high as 70\%. This enables us to investigate a more significant trend, but in real observation, the binary fractions of MW satellite galaxies can vary significantly. We thus show in Figure~\ref{fig:level3scale36} the result based on model-Ic, which has $f_\mathrm{binary}=$36\%. The main trends remain very similar between Figures~\ref{fig:level3scale70} and \ref{fig:level3scale36}, in the sense that $M(<r_\mathrm{half})$ are all inflated after including binary motions, whereas $M(200-300\mathrm{pc})$ tend to be mostly deflated. The amount of inflations in $M(<r_\mathrm{half})$ also slightly increases with the decrease in $\sigma_{v_r,\mathrm{eff}}$, though not as prominent as Figures~\ref{fig:level3scale70}. With decreased $f_\mathrm{binary}$, the amount of inflations becomes $\sim 10$\% at $\sigma_{v_r,\mathrm{eff}}$ of 3~km/s. The amount of deflations, on the other hand, does not show significant decrease, perhaps because the deflations are related to the gradient/shape in the velocity dispersion profile, instead of the absolute amount of inflations in the overall velocity dispersion.  

\subsection{model-II applied to scaled level-3 resolution} 
\subsubsection{Error free case to test the model dependence}

\begin{figure} 
\begin{center}
\includegraphics[width=0.49\textwidth]{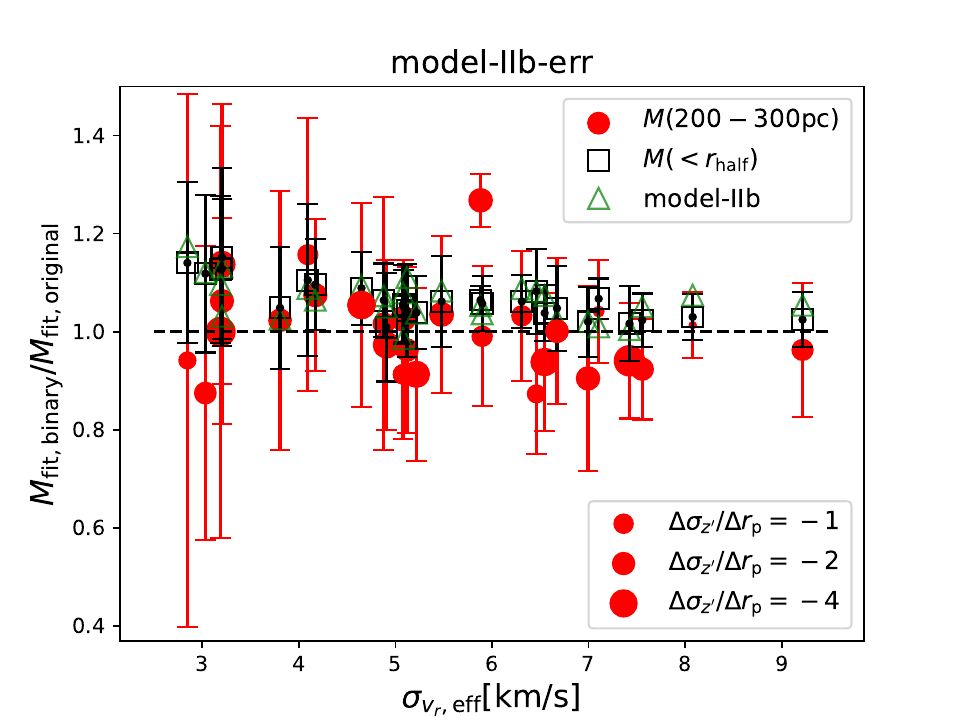}
\end{center}
\caption{Similar to the right plot of Figure~\ref{fig:level3scalemoe}, but observational errors of 3~km/s have been added to the line-of-sight velocities (model-IIb-err with 36\% of binary fraction). Note model-IIa-err with higher binary fraction of 70\% show very similar trends after including observational errors, so we choose not to repeatedly show the results. }
\label{fig:level3scalemoeerr}
\end{figure}

\begin{figure*} 
\begin{center}
\includegraphics[width=0.49\textwidth]{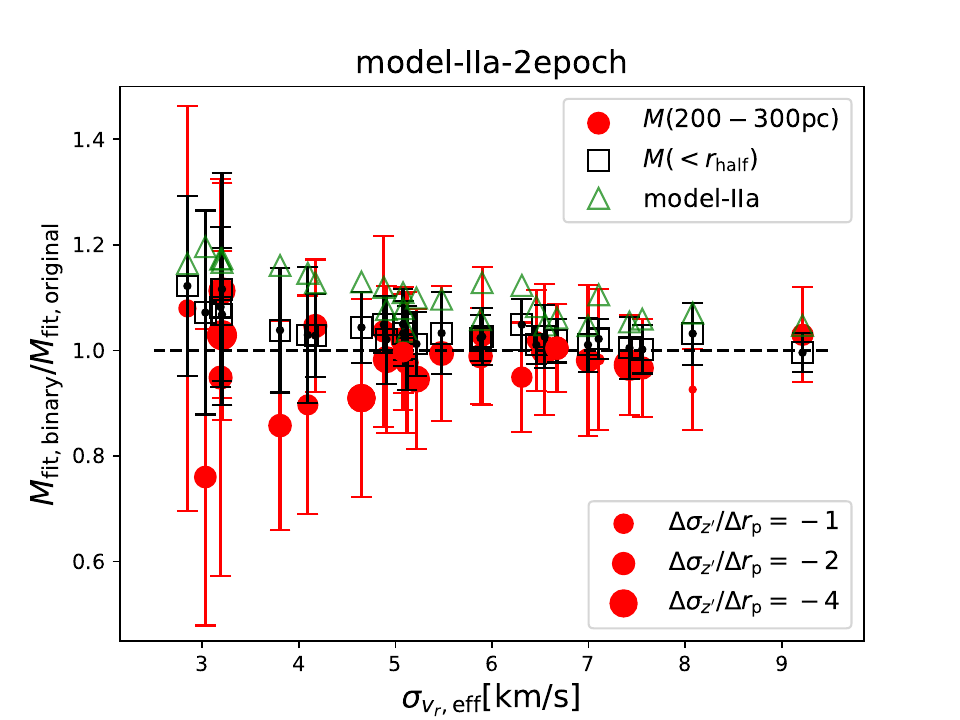}%
\includegraphics[width=0.49\textwidth]{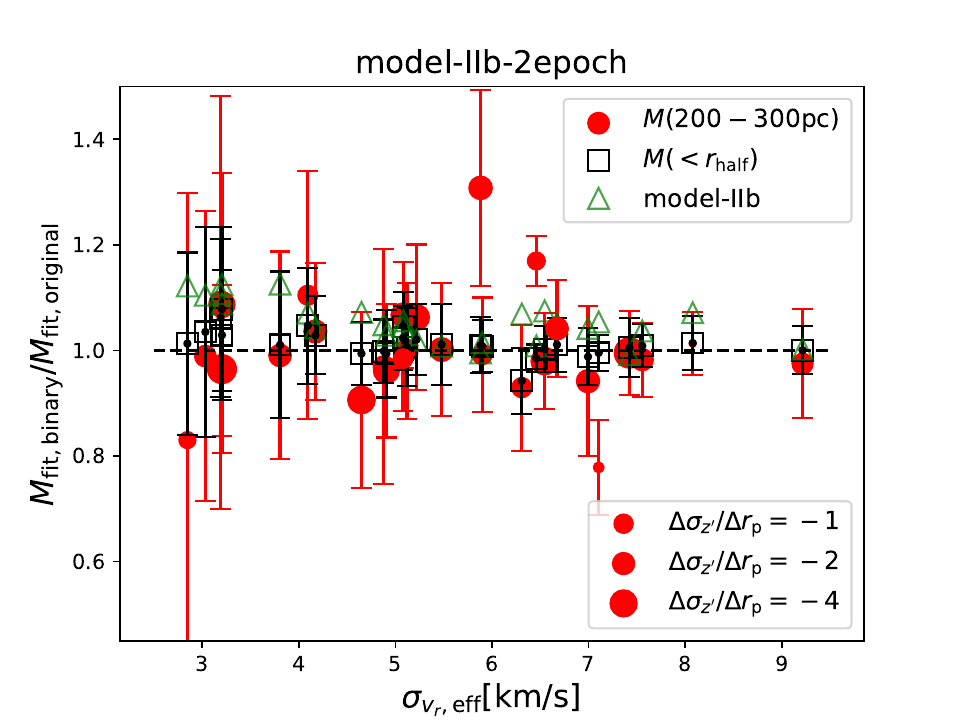}
\end{center}
\caption{{\bf Left:} Similar to the left plot of Figure~\ref{fig:level3scalemoe}, but after including observational errors of 3~km/s in line-of-sight velocities and after incorporating binary orbital motions, we drop tracers whose changes in their line-of-sight velocities are greater than 5~km/s in one year of time. This is based on model-IIa-2epoch with 70\% of binary fraction. {\bf Right:} Similar to the left plot, but is based on model-IIb-2epoch with 36\% of binary fraction. In both plots, the size of red dots is inversely proportional to the averaged radial gradient of the line-of-sight velocity dispersion profile over 0.2$r_\mathrm{half}$ and 2$r_\mathrm{half}$. The more negative the gradients are, the larger the symbol sizes.}
\label{fig:level3scalemoeerr2epoch}
\end{figure*}

With model-I, we have seen how binary motions affect the dynamical constraints in a radius dependent way, because the velocity dispersions of realistic dwarf galaxies can show negative radial gradients. However, model-I \citep{1991A&A...248..485D} is based on relatively old observations. In this subsection, we move on to consider model-II based on \cite{2017ApJS..230...15M}, which considers joint distributions of different orbital elements and is based on more recent observations. 

The results are shown in Figure~\ref{fig:level3scalemoe} for $f_\mathrm{binary}=$70\% and 36\%. Comparing the previous Figures~\ref{fig:level3scale70} and \ref{fig:level3scale36} based on model-I, we can see the main trends remain very similar, when $f_\mathrm{binary}$ is the same but the binary orbital element distribution models are different. Despite the differences in the models, $M(<r_\mathrm{half})$ are almost all inflated, and $M(200-300\mathrm{pc})$ are more likely deflated. Red dots below unity on average have more negative radial gradients in their LOSV dispersion profiles. The green triangles are repeats of the black squares from Figures~\ref{fig:level3scale70} and \ref{fig:level3scale36}, which are consistent with the new measurements, indicating no prominent model dependencies.

\subsubsection{Observational errors and multi-epoch data}

Now we consider more realistic cases after including observational errors and the mock of two-epoch observations with model-II. Figure~\ref{fig:level3scalemoeerr} is similar to the right plot of Figure~\ref{fig:level3scalemoe}. It is based on $f_\mathrm{binary}=$36\%, but we have included a typical error of 3~km/s to the LOSVs. We can see the black squares and green triangles almost overlap with each other, despite of the inclusion of observational errors. However, we also note in Figure~\ref{fig:level3scalemoeerr}, the red dots are slightly more symmetrically distributed around unity than the right plot of Figure~\ref{fig:level3scalemoe}, perhaps indicating the inclusion of observation errors can help to weaken the radial gradients in the LOSV dispersion profiles and the deflations in $M(200-300\mathrm{pc})$. 

Note here we only show the results after including observational errors for model-IIb-err, but not model-IIa-err. This is because all trends based on the comparisons between model-IIa and model-IIa-err are very similar, i.e., we see very similar amounts of inflations in $M(<r_\mathrm{half})$, and the deflations in $M(200-300\mathrm{pc})$ are slightly weakened. So we avoid repeatedly showing the results. 

For results in Figure~\ref{fig:level3scalemoeerr2epoch}, we further exclude star particles which have more than 5~km/s of changes in their LOSVs across one year of time. We create the LOSV for the second observation by adding one year of time and recalculate the LOSV at the new true anomaly ($f'$). This 5~km/s of threshold is applied to the LOSVs after including binary motions and observational errors of 3~km/s. Once we determine the star particles to be excluded, they are excluded from the tracer populations both before and after including binary motions, though the threshold itself is determined after incorporating binaries. 

In the left plot of Figure~\ref{fig:level3scalemoeerr2epoch}, the 70\% of binary fraction still leads to $\sim$10\% of inflations in $M(<r_\mathrm{half})$ at $\sigma_{v_r,\mathrm{eff}}\sim3$~km/s, which drops to $\sim1-3$\% at $4-8$~km/s. The deflations in $M(200-300\mathrm{pc})$ still exist, with more red dots below unity, but they become closer to unity, especially at $\sigma_{v_r,\mathrm{eff}}>5$~km/s. In the right plot, with the lower 36\% of binary fraction, there is almost no systematic bias at every $\sigma_{v_r,\mathrm{eff}}$. The black squares stay very close to zero. The red dots in the right plot still have large scatters but no longer show prominent biases towards below unity. Our results thus indicate that with a not extreme binary fraction, and typical observational errors of 3~km/s to the LOSVs, two epoch observations across one year of time are enough to avoid binary orbital motions affecting the dynamical modeling outcomes at 3~km/s$<\sigma_{v_r,\mathrm{eff}}<$9~km/s. 

\subsubsection{The 1 to 3~km/s region}

\begin{figure*} 
\begin{center}
\includegraphics[width=0.49\textwidth]{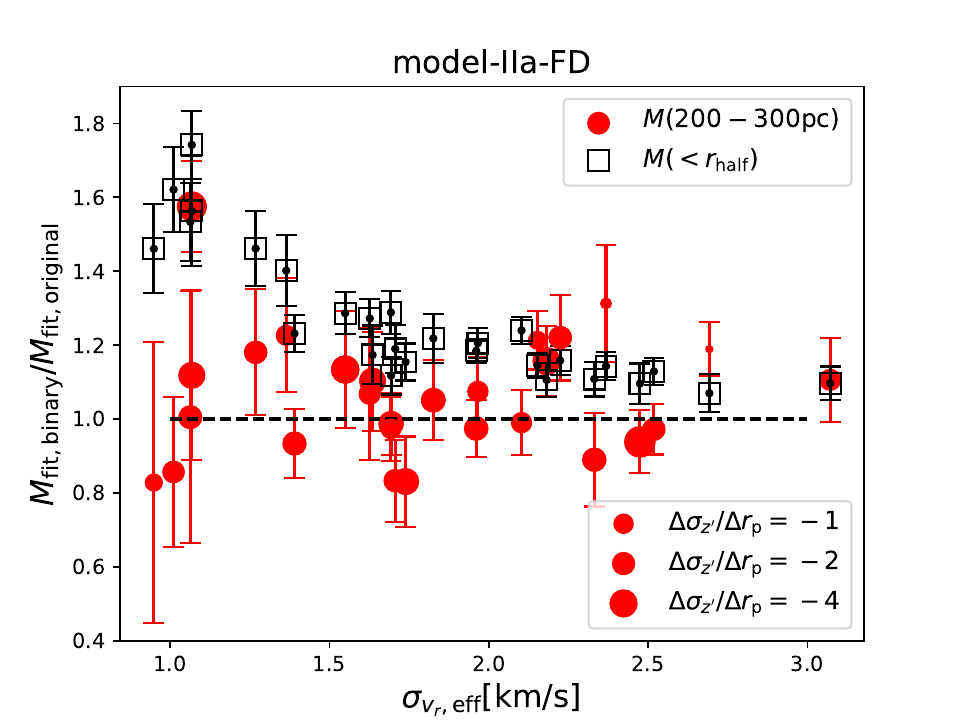}
\includegraphics[width=0.49\textwidth]{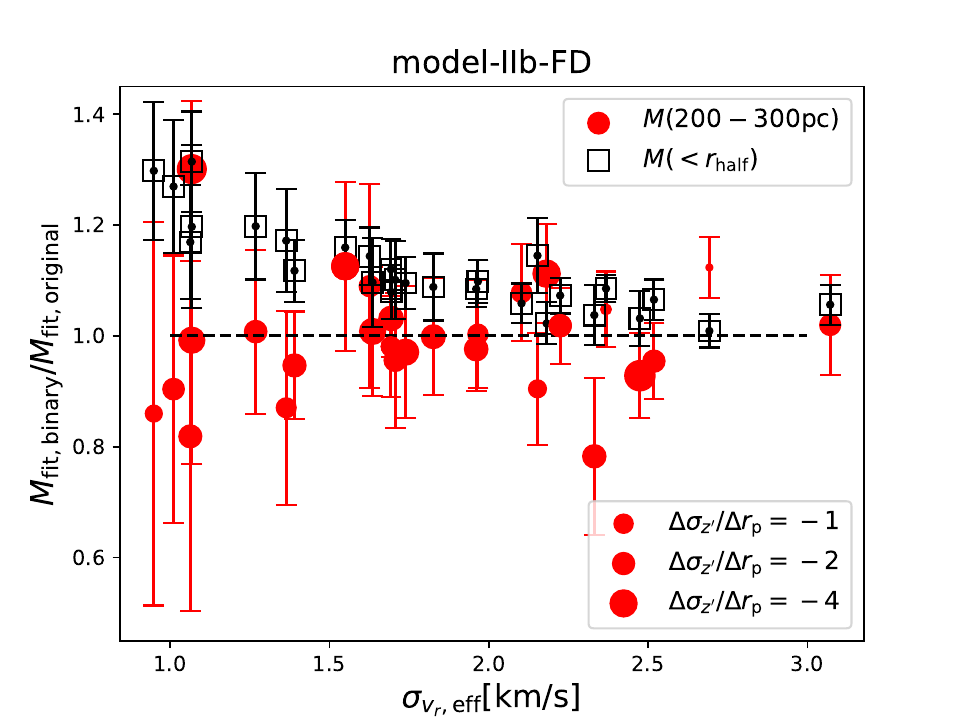}%
\end{center}
\caption{Similar to Figure~\ref{fig:level3scalemoeerr2epoch}, but is based on model-IIa-FD ($f_\mathrm{binary}=$70\%) and model-IIb-FD ($f_\mathrm{binary}=$36\%) to investigate how binary motions affect the 1 to 3~km/s region in line-of-sight velocity dispersions. The binary motions are scaled up by a factor of 10.5. Two epoch mock observations are adopted to exclude tracer stars with greater than 5~km/s of change in their line-of-sight velocities, after including observational error of 1~km/s. In both plots, the size of red dots is inversely proportional to the averaged radial gradients of the intrinsic line-of-sight velocity dispersion profiles over 0.2$r_\mathrm{half}$ and 2$r_\mathrm{half}$. The more negative the gradients are, the larger the symbol sizes.}
\label{fig:FD}
\end{figure*}

So far we have investigated how binaries affect the dynamical modeling outcomes at effective intrinsic velocity dispersions of 3~km/s$<\sigma_{v_r,\mathrm{eff}}<$9~km/s. We have tried different binary orbital element distribution models, cases with or without observational errors and mock of multi-epoch observations. In the current subsection we move on to investigate the region of 1~km/s$<\sigma_{v_r,\mathrm{eff}}<$3~km/s. Here we scale up the binary orbital motions by a factor of 10.5, and apply \textsc{jam} to dwarf galaxies with $\log_{10}M_\ast/\msun>7.5$ from \textsc{auriga} level-3 simulations. With the scaling of 10.5, we are equivalently investigating dwarfs with 1~km/s$<\sigma_{v_r,\mathrm{eff}}<$3~km/s.

The results are shown in Figure~\ref{fig:FD}. We have excluded tracer stars which have changes in their LOSVs greater than 5~km/s across two observations separated by one year, subject to an observational error of 1~km/s in LOSVs. With $f_\mathrm{binary}=$70\% and 36\%, $M(<r_\mathrm{half})$ can still be significantly inflated by $\sim$60\% and 30\% at $\sigma_{v_r,\mathrm{eff}}\sim 1$~km/s. Over 1.5 to 3~km/s, the inflations in $M(<r_\mathrm{half})$ range from $\sim$28\% to 10\% for the binary fraction of 70\%. With the lower binary fraction of 36\%, the amount of inflations in $M(<r_\mathrm{half})$ is close to zero at 3~km/s, but reaches $\sim$15\% to 5\% at 1.5 to 2.5~km/s.

In comparison, $M(200-300\mathrm{pc})$ are not prominently deflated in the left plot of Figure~\ref{fig:FD}. The red dots distribute more symmetrically around the black horizontal dashed line in the left plot of Figure~\ref{fig:FD}. Unlike previous figures, we do not see more red dots below unity. This is perhaps because with a high binary fraction of 70\% and over the range of 1~km/s$<\sigma_{v_r,\mathrm{eff}}<$3~km/s, the fractional inflations in the velocity dispersions due to binary motions are very strong at every radii, and the radial gradients in the intrinsic velocity dispersions are no longer significant compared with the additional dispersions introduced by binary motions, and thus we do not see obvious deflations in $M(200-300\mathrm{pc})$. On the other hand and in the right plot of Figure~\ref{fig:FD}, where $f_\mathrm{binary}$ gets lowered to 36\%, we can still see the trend that there are more red dots below unity. 

We thus conclude that for ultra faint dwarf galaxies whose intrinsic velocity dispersions are close to 1~km/s, they are more significantly affected by the existence of binaries. This is true even for a mild binary fraction of 36\% and after excluding tracer stars with prominent changes in the LOSVs with two epoch of observations. 

In the end, we note that in real observation, the number of observed stellar tracers can be very small (from a few to a few tens only) for such ultra faint dwarfs with 1~km/s$<\sigma_{v_r,\mathrm{eff}}<$3~km/s. In our analysis, we adopt a large sample of tracers, which do not fully represent the real case. So our results in this subsection can only be regarded as an ensemble averaged behavior for a large sample of such ultra faint dwarfs. In real observation, the dynamical constraints for dwarf galaxies with 1~km/s$<\sigma_{v_r,\mathrm{eff}}<$3~km/s can have large system-to-system scatters, and are expected to have large statistical errors depending on the tracer sample size.

\section{Discussions and conclusions}
\label{sec:concl}

In this study, we investigate how binary orbital motions affect the dynamical modeling outcomes of dwarf galaxies, by using realistic tracers constructed from star particles for 17 and 28 dwarf galaxies from the \textsc{auriga} level-2 and level-3 suites of simulations. 

Level-2 resolution only has one MW-like system, but it can have $\sim$40 star particles for dwarf galaxies with $M_{*} \sim 3\times10^4\msun$, which allows direct and initial checks for the effect of binary motions. However, it still ends up with large statistical errors, and the internal dynamics of dwarf systems cannot be well resolved with a few tens of particles. Thus we also select dwarf galaxies more massive than $10^{7.5}\msun$ from six MW-like systems of the level-3 resolution. We scale up the level of binary orbital motions by factors of 3.5 or 10.5, which is equivalent to decreasing the velocity dispersions of the dwarf galaxies, but at the same time we can have enough number of tracer particles to ensure good statistics. 

On the basis of the above mock dwarf systems and their tracer star samples, we incorporate binary orbital motions by sampling the orbital element distributions of binary systems based on observations of solar neighborhood stars \citep{1991A&A...248..485D,2017ApJS..230...15M}. We further apply the Jeans Anisotropic Multi-Gaussian Expansion method (\textsc{jam}) to investigate how the best-constrained dynamical mass is changed before and after incorporating the binaries. 

For level-2 resolution and by sampling binary orbital motions from the \cite{1991A&A...248..485D} model with a 70\% fraction of binaries, the amount of inflations in the best constrained dynamical mass prominently increases with the decrease in the LOSV dispersions ($\sigma_{v_r}$) of dwarf galaxies. 

With level-3 resolution, we sample binary orbital motions from both the \cite{1991A&A...248..485D} model and the more recent \cite{2017ApJS..230...15M} model. The latter study has explicitly considered joint distributions of different orbital elements based on more recent observations. We find the amount of inflations/deflations and the overall trends with the above two orbital motion models are statistical consistent each other with the same binary fraction, indicating no prominent dependencies on binary orbital element distribution models. 

Our major results are based on the \cite{1991A&A...248..485D} model, we find the total masses within the half-mass radius of tracers, $M(<r_\mathrm{half})$, are all inflated after including binaries, reaching maximums of 15\% for 70\% of binary fraction ($f_\mathrm{binary}$) at effective LOSV dispersion of $\sigma_{v_r,\mathrm{eff}}\sim$3~km/s, and decreases to 10\% for $f_\mathrm{binary}=$36\%. 

Interestingly, the dynamically constrained mass in central regions tend to be deflated after including binary motions. In particular, when the inflations in $M(<r_\mathrm{half})$ increases with the decrease in $\sigma_{v_r,\mathrm{eff}}$, the deflations in the central density do not show clear dependencies on $\sigma_{v_r,\mathrm{eff}}$. Dwarf galaxies with $\sigma_{v_r,\mathrm{eff}}$ in the range of 3-8~km/s can have their dynamically constrained central density deflated by up to 10-20\% due to binary motions. Besides, such deflations do not show prominent dependences on $f_\mathrm{binary}$ either. 

The deflations in the central density are due to the negative radial gradient in the velocity dispersion profiles. The velocity dispersion significantly decreases with the increase in radius. As a result, the LOSV dispersions are much smaller in outskirts, bringing in much more significant fractional increases at larger radius, and thus more significant inflations in the best constrained dynamical masses in outer regions. The fractional increase is much less significant in central regions. Thus the dynamical masses in central regions are inflated less. Due to a reconcile to better fit the velocity map in outskirts, the best constrained dynamical masses in inner regions are often deflated. 

The deflation in the central density is important, because most previous studies based on Monte Carlo sampling of binary motions do not include radial gradients in their intrinsic velocity dispersions. Since binary motions inflate the intrinsic velocity dispersions, a general impression is held that the dynamically constrained mass of ultra faint dwarf galaxies are inflated due to binary motions. We show in this study, for the first time, that the effect of binary motions on the dynamical mass constraints is radius dependent. In inner regions, deflations are more likely to happen. 

Moreover, since it is the total mass in inner regions of dwarf galaxies which are more sensitive to the inner density slopes, the deflations in the central density can be more closely related to the core-cusp problem. Since binary orbital motions can deflate the inner dynamical mass, the inner density slopes can be under-estimated. As a result, cuspy dwarf galaxies might be determined to be biased to cored if they have strong negative radial gradients in their velocity dispersion profiles. In fact, we have shown in Paper I that global contractions of the dwarf galaxies can result in under-estimated inner density profiles for steady state models. In particular, we have shown that for a few Sagittarius dwarf Spheroid like systems, the \textsc{jam} constrained inner density profiles are significantly more flattened than the truth, on the basis of which, the deflated central densities due to binaries can further make the best constrained inner densities more flattened.

After including a 3~km/s of observational error to the LOSVs, the trends remain very similar, but the deflations in the central density seem to be slightly weakened. By further discarding tracer star particles which have greater than 5~km/s changes in their LOSVs after including binary motions and 3~km/s of observational errors, we find the inflations decrease to almost zero for none extreme binary fractions of 36\%. Our results thus indicate that for none extreme binary fractions of 30-40\%, and with typical observational errors to the LOSV and multi-epoch data, binary orbital motions are unlikely to significantly affect the dynamical modeling outcome at $\sigma_{v_r,\mathrm{eff}}\sim3-9$~km/s.  On the other hand, for more extreme binary fractions of 70\%, there might still be $<\sim10$\% of inflations in $M(<r_\mathrm{half})$ at $\sigma_{v_r,\mathrm{eff}}\sim3$~km/s, which is, however, not statistically significant compared with the errors in our analysis.

In the end, we investigate the region of 1~km/s$<\sigma_{v_r,\mathrm{eff}}<3$~km/s. We find even with two epoch observations crossing one year of time to exclude stars whose changes in LOSVs are greater than 5km/s, $M(<r_\mathrm{half})$ can still be significantly inflated by $\sim$60\% and 30\% at $\sigma_{v_r,\mathrm{eff}}\sim1$~km/s, for binary fractions of 70\% and 36\%, respectively. At 1.4~km/s$<\sigma_{v_r,\mathrm{eff}}<3$~km/s, the inflations in $M(<r_\mathrm{half})$ range from $\sim$28\% to 10\% with binary fraction of 70\%. With the binary fraction of 36\%, the inflation in $M(<r_\mathrm{half})$ is close to zero at 3~km/s, but reaches $\sim$15\% to 5\% at 1.5 to 2.5~km/s.

\acknowledgments
This work is supported by NSFC (12022307, 12273021, 12203100), the China Manned Space (CSST) 
Project with No. CMS-CSST-2021-A02 and CMS-CSST-2021-A08, the National Key Basic Research and 
Development Program of China (No. 2018YFA0404504), 111 project (No. B20019) and Shanghai Natural 
Science Foundation (No. 19ZR1466800). We thank the sponsorship from Yangyang Development Fund. 
WW is grateful for discussions with Chao Liu and Haifeng Wang on binary orbital element distribution, 
with Kaiming Cui on calculation of stellar parameters and with Haijun Tian on binary physics. We thank 
the invitation and host by Sarah Bird and Haijun Tian during the 2022 Gaia Sprint event at the Three 
Georges University. The computation of this work is carried out on the \textsc{Gravity} 
supercomputer at the Department of Astronomy, Shanghai Jiao Tong University, and is partly supported 
by the STFC DiRAC HPC Facility, at the Institute of Computational Cosmology (ICC), Durham University. 
LZ acknowledges funding from CAS Project for Young Scientists in Basic Research, Grant No. YSBR-062, 
and National Natural Science Foundation of China under grant No. Y945271001. 
RG acknowledges financial 
support from the Spanish Ministry of Science and Innovation (MICINN) through the Spanish State Research 
Agency, under the Severo Ochoa Program 2020-2023 (CEX2019-000920-S), and support from an STFC Ernest 
Rutherford Fellowship (ST/W003643/1). 
ZZL acknowledge the support by ISF grants 861/20, 3061/21, 
DFG/DIP grant STE1869/2-1 GE625/17-1 and MSCA Fellowship (101109759).


\bibliography{master}{}
\bibliographystyle{aasjournal}




\end{document}